\documentclass[]{article}
\usepackage[margin=2.5cm]{geometry}
\usepackage{graphicx}
\usepackage[utf8]{inputenc}
\usepackage{epstopdf, epsfig}
\usepackage[hidelinks]{hyperref}
\usepackage{authblk}

\usepackage{amsmath,amssymb}
\usepackage{graphicx,import,subcaption}
\usepackage[]{natbib}
\setcitestyle{semicolon,open={(},close={)}}
\usepackage{doi}
\usepackage{color}
\usepackage{tikz}
\usetikzlibrary{calc}
\usetikzlibrary{backgrounds}
\usepackage{standalone}
\usepackage{pgfplots}
\pgfplotsset{compat=1.17}

\title{Numerical simulations of viscous fingering in fractured porous media}

\author[1,2]{Runar L. Berge\thanks{Corresponding Author: runar.lie.berge@hivolda.no}}
\author[2]{Inga Berre}
\author[2]{Eirik Keilegavlen}
\author[2]{Jan M. Nordbotten}
\affil[1]{Department of Science, Volda University College, Volda, Norway}
\affil[2]{Department of Mathematics, University of Bergen,
Bergen, Norway}

\renewcommand\d[0]{\ensuremath{\operatorname{d}\!}}

\renewcommand\vec[0]{\ensuremath{\boldsymbol}}
\newcommand\Pen{\mbox{\textit{Pe}}}

\def\legendWidth{0.13\textwidth}

\begin{document}
\maketitle

\begin{abstract}%
  The effect of heterogeneities induced by highly permeable fracture networks on
  viscous miscible fingering in porous media is examined using high-resolution
  numerical simulations. We consider the planar injection of a less viscous fluid
  into a two-dimensional fractured porous medium which is saturated with a more
  viscous fluid. This problem contains two sets of fundamentally different
  preferential flow regimes; the first is caused by the viscous fingering and the
  second is due to the permeability contrasts between the fractures and the rock
  matrix. We study the transition from the regime where the flow is dominated by the
  viscous instabilities, to the regime where the heterogeneities induced by the
  fractures define the flow paths. Our findings reveal that even minor permeability 
  differences between the rock matrix and fractures significantly influence the 
  behavior of viscous fingering. The interplay between the viscosity contrast and 
  permeability contrast leads to the preferential channeling of the less viscous 
  fluid through the fractures.
  Consequently, this channeling process stabilizes the displacement front within the rock matrix, ultimately suppressing the occurrence of viscous fingering, particularly for higher permeability contrasts. We explore three fracture geometries; two 
  structured and one random configuration, and identify a complex interaction between 
  these geometries and the development of unstable flow. While we find that the most 
  important factor determining the effect of
  the fracture network is the ratio of fluid volume flowing through the fractures and 
  the rock matrix, the exact point for the cross-over regime is dependent on the 
  geometry of the fracture network.
\end{abstract}

\section{Introduction}
Viscous fingering in a porous medium is a process that occurs within a
wide range of flow processes, including CO$_2$ storage, enhanced oil recovery,
and geothermal energy systems. The porous rocks that are relevant for these
applications have often undergone fracturing processes, and fractures generally extend throughout the entire reservoir~\citep{berkowitz2002characterizing}. These
fractures impose a structural constraint on fluid flow and transport through the
domain, and this paper targets the interplay between unstable miscible fingering
and channeling through fractures that are more permeable than the surrounding rock
matrix.

When a less viscous fluid displaces a more viscous fluid, hydrodynamical
instabilities are induced that may cause viscous
fingering~\citep{homsy1987viscous}. Miscible viscous fingering in a homogeneous
medium has been extensively studied using a variety of different methods, including
linear stability analysis~\citep{tan1986stability,tan1987stability}, numerical
simulations~\citep{nijjer_2018,zimmermann1991nonlinear,zimmerman1992viscous,christie1989high-resolution},
and laboratory
experiments~\citep{kopf-still1988nonlinear,baxri1992miscible,petitjeans1999miscible,chui2015interface}. These
studies have characterized the evolution of the viscous fingering from the
initialization to the later stages. The onset of viscous fingers can be predicted
by the wave number with the highest growth rate from linear stability analysis. As
the instabilities grow, the flow is governed by different mechanisms such as
splitting, merging, or shielding~\citep{homsy1987viscous}. The exact behavior of
the displacement front is determined by the parameter regime, and the viscous
instabilities depend on a wide range of factors, including
gravity~\citep{manickam1995fingering}, miscibility~\citep{fu2017viscous},
anisotropic dispersion~\citep{yortsos1988dispersion},
mixing~\citep{jha2011quantifying}, and reactions~\citep{wit2004miscible}.

Porous media found in geological formations are in general not homogeneous, but
vary on a wide range of length scales, from the pore scale to the reservoir
scale. \citet{wit1997viscous,wit1997viscous_b} and~\citet{nicolaides2015impact}
considered different randomly varying permeability fields and found that the
preferential flow paths given by the permeability field are competing with the
unstable flow paths from the viscous fingering. More structured permeability
fields have also been studied, with the main focus on layered porous media. When
the flow is predominantly aligned with the permeable layers, the heterogeneities
tend to cause channeling through the
domain~\citep{nijjer2019stable,sajjadi2013scaling,shahnazari2018linear}.

When fractures are present in a porous medium, the fractures can act as
preferential flow paths due to high permeability contrasts with the rock
matrix. Fractures can give rise to complicated flow and transport patterns through
a porous medium, and they present challenges to modeling~\citep{neuman2005trends},
upscaling~\citep{nissen2018heterogenity}, and
discretization~\citep{flemish2018benchmarks}, which are further complicated by
fracture geometries~\citep{berge2019}. These challenges have received considerable
attention the last two decades; for an overview, we refer the reader to the
textbooks by~\citet{dietrich2005flow} and~\citet{sahimi2011flow}, and the review
study by~\citet{berre2018flow} on different mathematical models of fracture flow. Despite the extensive work on both viscous fingering and flow in fractures, little
attention has been given to characterize the interaction between the viscous
instabilities and preferential flow paths through a fractured porous
material. \citet{budeck2015} show experimental and numerical results for viscous
fingering in microchannels that behave similar to fracture networks, however, they do not consider flow in the material between the channels that would represent the rock matrix. \citet{zhang1998} include fluid flow in both fractures and
rock matrix in their numerical experiments of radial flow, and investigate how the viscous fingering in immiscible displacement is affected by various fracture network parameters. Fractures can be represented in models in various ways, however, when studying the dynamics between fractures and viscous fingering it warrants an explicit representation of fractures to capture the intricate effects of fracture geometry. A popular class of conceptual models for fractured porous medium that explicitly represents fractures in the domain are the
discrete-fracture-matrix (DFM) models. An efficient approach for representing DFM models is to consider fractures as lower-dimensional inclusions in the rock matrix due to their large aspect ratios~\citep{martin2005,karimi-fard2003efficient}. This
approach has lead to the development of accompanying discretization strategies;
see, e.g., \citep{flemish2018benchmarks,nordbotten2019unified,reichenberger2006}.

The current paper considers miscible flow in
both the rock matrix and fractures, and addresses three key questions: (i) what factors related to the fracture flow are most important when studying the effect on the viscous instabilities, (ii) how does the interplay between the viscous
instabilities and fractures affect the flow patters in the rock matrix, and (iii)
how does different fracture geometries affect the answer to question (i) and
(ii). To address these questions, we exploit recent advancements within modeling
and simulation tools that focus on flow and transport in fractured porous
media. We employ a DFM model, and by state-of-the-art numerical simulations
characterize unstable miscible displacement through various fracture network
geometries. This modelling approach is necessitated by the high and discontinuous permeability contrasts between fractures and rock matrix as well as the complexity of the random fracture networks investigated in this study. Furthermore, to obtain simulations that extend up to the point of viscous fingering shutdown, we employ adaptive strategies. It is worth noting that the intricacies of implementing these adaptive strategies within the framework of DFM models can be challenging. As a result, we make our implementation open source to facilitate further research and exploration in this field.

The remaining manuscript is laid out as follows. In
Section~\ref{sec:governing_equations}, we present the governing equations for
miscible viscous flow in a fractured porous media. Special attention is given to
the mixed-dimensional DFM model. In Section~\ref{sec:scaling}, the equations are
represented in dimensionless form, and we discuss the three new dimensionless
numbers that appear due to the fractures, and in Section~\ref{sec:numerical_method},
we discuss the numerical method. The results given in
Section~\ref{sec:hom_and_layered} are divided into three subsection, one for each fracture
geometry considered, and finally, we give concluding remarks.

\section{Governing equations}\label{sec:governing_equations}
\begin{figure}
  \begin{subfigure}[]{0.7\textwidth}
    \centering%
    \def\svgwidth{\textwidth}%
    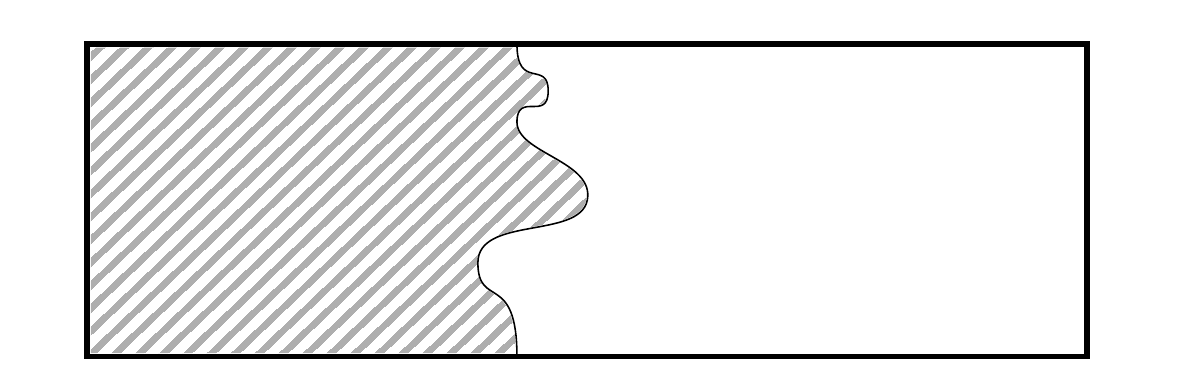
    \caption{}
  \end{subfigure}%
  \hfill
  \begin{subfigure}[]{0.2\textwidth}
    \centering%
    \def\svgwidth{\textwidth}%
\begingroup%
  \makeatletter%
  \providecommand\color[2][]{%
    \errmessage{(Inkscape) Color is used for the text in Inkscape, but the package 'color.sty' is not loaded}%
    \renewcommand\color[2][]{}%
  }%
  \providecommand\transparent[1]{%
    \errmessage{(Inkscape) Transparency is used (non-zero) for the text in Inkscape, but the package 'transparent.sty' is not loaded}%
    \renewcommand\transparent[1]{}%
  }%
  \providecommand\rotatebox[2]{#2}%
  \newcommand*\fsize{\dimexpr\f@size pt\relax}%
  \newcommand*\lineheight[1]{\fontsize{\fsize}{#1\fsize}\selectfont}%
  \ifx\svgwidth\undefined%
    \setlength{\unitlength}{307.3660807bp}%
    \ifx\svgscale\undefined%
      \relax%
    \else%
      \setlength{\unitlength}{\unitlength * \real{\svgscale}}%
    \fi%
  \else%
    \setlength{\unitlength}{\svgwidth}%
  \fi%
  \global\let\svgwidth\undefined%
  \global\let\svgscale\undefined%
  \makeatother%
  \begin{picture}(1,1.22650006)%
    \lineheight{1}%
    \setlength\tabcolsep{0pt}%
    \put(0,0){\includegraphics[width=\unitlength,page=1]{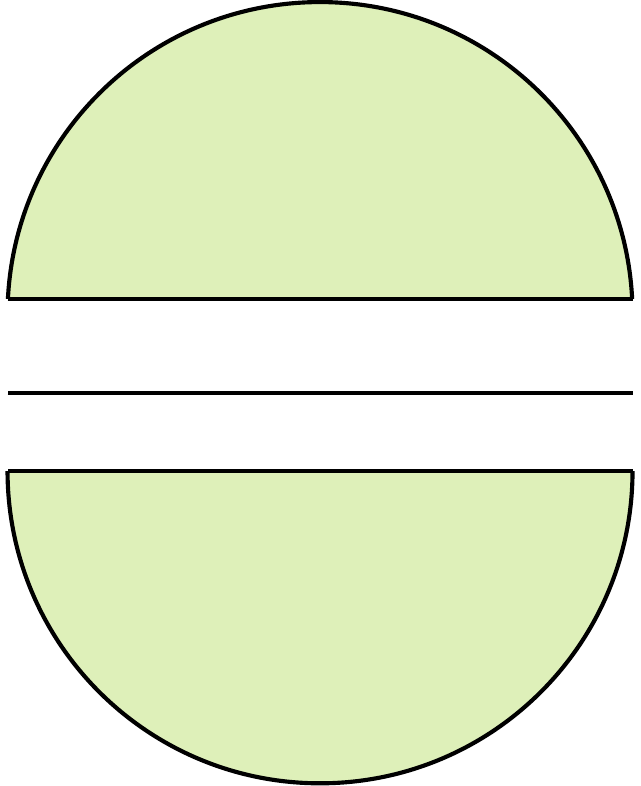}}%
    \put(0.51571699,0.6452036){\color[rgb]{0,0,0}\makebox(0,0)[t]{\lineheight{0}\smash{\begin{tabular}[t]{c}$\Omega_f$\end{tabular}}}}%
    \put(0.51571699,0.79432977){\color[rgb]{0,0,0}\makebox(0,0)[t]{\lineheight{0}\smash{\begin{tabular}[t]{c}$\Gamma^+$\end{tabular}}}}%
    \put(0.51571699,0.24343198){\color[rgb]{0,0,0}\makebox(0,0)[t]{\lineheight{0}\smash{\begin{tabular}[t]{c}$\Omega_m$\end{tabular}}}}%
    \put(0.51571699,0.99637312){\color[rgb]{0,0,0}\makebox(0,0)[t]{\lineheight{0}\smash{\begin{tabular}[t]{c}$\Omega_m$\end{tabular}}}}%
    \put(0.51571699,0.39624704){\color[rgb]{0,0,0}\makebox(0,0)[t]{\lineheight{0}\smash{\begin{tabular}[t]{c}$\Gamma^-$\end{tabular}}}}%
    \put(0,0){\includegraphics[width=\unitlength,page=2]{interface_notation.pdf}}%
    \put(0.77309383,0.51321013){\color[rgb]{0,0,0}\makebox(0,0)[lt]{\lineheight{0}\smash{\begin{tabular}[t]{l}$\vec\nu^-$\end{tabular}}}}%
  \end{picture}%
\endgroup%

    \caption{\label{fig:interface_notation}}
  \end{subfigure}
  \caption{A schematic representation of the problem setup. (a) A domain of length $L$ and width $H$. The rock matrix is denoted by
    $\Omega_m$ and contains two fractures denoted by $\Omega_f$. The transversal distance between
    the fractures is given by $W$. The porous medium is initially filled by a
    fluid of viscosity $\mu_d$ and displaced by a fluid of viscosity $\mu_u$. (b)
    Conceptual figure of a fracture. The rock matrix, $\Omega_m$, has two
    boundaries, $\Gamma^+$ and $\Gamma^-$, associated with the fracture,
    $\Omega_f$. Note that $\Gamma^+$, $\Gamma^-$ and $\Omega_f$ all coincide
    geometrically, but are separated in the illustration for a better visualization.}
  \label{fig:fractured_domain}
\end{figure}
A schematic illustration of the problem setup is shown in
Figure~\ref{fig:fractured_domain}. We consider two incompressible fluids with
different viscosities $\mu_u \le \mu_d$, in a periodic fractured porous media. The
two fluids are completely miscible and can therefore be modeled as a single-phase
fluid with two components. The fractures, denoted by $\Omega_f$, are considered as
one-dimensional (1D) inclusions, embedded in a surrounding two-dimensional (2D)
rock matrix denoted by $\Omega_m$. Throughout this paper we use subscripts ``$f$''
and ``$m$'' when it is necessary to distinguish fracture and rock matrix values,
respectively. The intersection between the rock matrix domain boundary and
the fractures defines one interface on each side of the fracture, which we call the
positive and negative interface,
$\partial\Omega_m\cap\Omega_f = \{\Gamma^+,\Gamma^-\}$. We denote the unit normal
vector on $\Gamma^\pm$ that points outward from $\Omega_m$ by $\vec \nu^\pm$, as
depicted in Figure~\ref{fig:interface_notation}.

The viscosity of the mixed fluid is assumed to depend exponentially on the mass ratio, $c$, of
the two components, $\mu = \mu_de^{-Rc}$, with $R = \ln \mu_d / \mu_u$, and the
fluid follows Darcy's law both in the rock matrix and in the fractures. The rock
matrix has a constant scalar permeability $k_m$, the porosity $\phi_m$ is
constant, and the diffusivity, $D_m$, is assumed isotropic and independent of the
component concentration. Thus, the governing equations in the rock matrix domain
$\Omega_m$ are given by
\begin{equation}
  \label{eq:matrix_equation}
  \begin{aligned}
    \vec u_m = -\frac{k_m}{\mu(c)}\nabla p_m, \quad \nabla \cdot \vec u_m = 0 \\
     \frac{\partial \phi_mc_m}{\partial t} + \nabla \cdot (c_m\vec u_m - D_m\nabla
    c_m) = 0
  \end{aligned} \qquad \text{in}\ \Omega_m,
\end{equation}
in addition to the boundary conditions. The first line of
Equation~\eqref{eq:matrix_equation} consists of Darcy's law, and mass conservation for an
incompressible fluid. The second line describes the advective and diffusive mass
transport of the components.

\subsection{Fracture flow}
Recall that the governing equations are written for some intermediate scale, large
compared to the pore size and fracture aperture, but small compared to the
extension of the fractures. This motivates a reduction of dimensionality where the
fractures are represented as lower-dimensional. For details on the derivation of
the reduced dimensional models we refer the reader
to~\cite{martin2005,flemish2018benchmarks,schwench2015}.

The fluid flow in the fractures is governed by Darcy's law and mass conservation. Assuming a lower-dimensional representation of the fractures, it is natural to consider flow rate through the fracture per unit length. This flow rate tangential to the fractures is then given by
\begin{equation}
  \label{eq:fracture_equation}
  \vec u_f = -\frac{a k_f}{\mu(c_f)}\nabla_{||} p_f \quad \text{in}\ \Omega_f,
\end{equation}
where $\nabla_{||}p_f$ is the pressure gradient in the tangential direction of the
fracture, $a$ is the fracture aperture, and $k_f$ is the fracture permeability. The value $a k_f$ is the effective fracture permeability, and this structure of the effective permeability is valid for fractures that have an aperture $a$ and are filled by a porous material (e.g, gouge) with permeability $k_f$. However, it can also be associated to the permeability of open fractures through the cubic law (see, e.g., \cite{jaeger2007fundamentals}).

Mass conservation in the fractures can be formulated as
\begin{equation}\label{eq:fluid_mass_conservation}
  \nabla_{||} \cdot \vec u_f - [\![\lambda]\!] = 0 \quad \text{in}\ \Omega_f,
\end{equation}
where $\nabla_{||}$ is the del-operator in tangential direction, and the term
$[\![\lambda ]\!] = \lambda^+ + \lambda^-$ is the sum of the fluid fluxes flowing from the rock matrix to fracture, defined by a Darcy type relation (see, e.g., \citep{martin2005}):
\begin{equation}\label{eq:advective_coupling}
  \vec u_m^\pm\cdot \vec \nu^\pm = \lambda^\pm =-\frac{k_f}{\mu(c^\pm)}\frac{(p_f - \text{tr}^\pm(p_m))}{a/2}
  \quad \text{on}\ \Gamma^\pm,
\end{equation}
where $\text{tr}^\pm(\cdot)$ is the trace operator restricted to the positive or
negative interface and $c^\pm$ is the upwind value on the interface:
\begin{equation}
    c^\pm = \begin{cases}
    \text{tr}^\pm(c_m) &\text{if}\ \lambda > 0\\
    c_f &\text{if}\ \lambda \le 0.
    \end{cases}
\end{equation}

The transport equation of the concentration in the fractures, $c_f$, is derived in
a similar manner as the pressure equation. We denote the sum of the advective and diffusive
flux from the rock matrix domain by $\lambda_c$, and write the conservation
equation in the fracture as
{\par\nobreak\noindent}
\begin{equation*}
  \frac{\partial a\phi_f c_f}{\partial t} + \nabla_{||} \cdot (c_f\vec u_f -
  D_fa\nabla_{||} c_f) - [\![\lambda_c]\!] = 0
  \quad \text{in}\ \Omega_f,
\end{equation*}
where $\phi_f$ and $D_f$ are the fracture porosity and diffusivity, respectively.
The total flux between the domains is given by
\begin{equation}\label{eq:coupling_flux}
  \lambda_c^\pm = c^\pm\lambda^\pm -\frac{D_f}{a/2}(c_f - \text{tr}^\pm(c_m))
  \quad \text{on}\ \Gamma^\pm.
\end{equation}
{\par\nobreak\noindent}

The intersection of two or more fractures is a 0D point where pressure continuity
and flux balance is enforced for all fractures that intersect in this
point. For intersecting fractures that have equal permeability, this is a
reasonable choice~\citep{stefansson2018}.

\subsection{Boundary and initial conditions}\label{sec:boundary_initial}
We consider flow that is periodic in the $y$-direction:
{\par\nobreak\noindent}
\begin{equation*}
  \begin{aligned}
    p(x, 0, t) &= p(x, H, t), & \vec u(x, 0, t) &= \vec u(x, H, t),\\
    c(x, 0, t) &= c(x, H, t), & D\nabla c(x, 0, t) &= D\nabla c(x, H,
    t).
  \end{aligned}
\end{equation*}
A constant flux, $U_m$, is enforced at the upstream boundary of
the rock matrix:
\begin{equation}\label{eq:upstream_bc}
  \vec u_m(X_u, y, t)\cdot\vec \nu = -U_m, \quad \ (X_u, y)\in \partial\Omega_m,
\end{equation}
where $X_u$ is the $x$-position of the upstream boundary and $\vec \nu$ is the
outer normal vector. The upstream flux on the fracture boundary,
\begin{equation}\label{eq:upstream_bc_f}
  \vec u_f(0, y, t)\cdot \vec \nu = -U_f = -\frac{k_fa}{k_m}U_m, \quad (X_u, y)\in \partial\Omega_f,
\end{equation}
is chosen such that the initial pressure drop for the test case with parallel
fractures only varies in the $x$-direction. A fixed pressure is given at the
downstream boundary that is located at $x$-coordinate $X_d$:
{\par\nobreak\noindent}
\begin{equation*}
  p_\alpha(X_d, y, t) = 0, \quad [X_d, y]\in \partial\Omega_\alpha,\quad \alpha\in\{m,f\}.
\end{equation*}

The initial condition is given by an almost sharp interface and a small
perturbation centered at $x=0$,
{\par\nobreak\noindent}
\begin{equation*}
  c_\alpha(x, y, t=0) = \frac{1}{2}+\frac{1}{2}\text{erf}\left(-\frac{x}{\sqrt{t_0}}\right)+U(x,y)\exp\left(-\frac{x^2}{t_0}\right),\quad \alpha\in\{m, f\}
\end{equation*}
where $\text{erf}(\cdot)$ is the error function, $U(x,y)\in[0,10^{-4}]$ is a
uniformly random perturbation added to initialize the viscous instabilities, and
$t_0=10^{-5}$ is chosen to aid the numerical scheme at early times. It is well
known that the onset time of the nonlinear viscous fingering is sensitive to
the amplitude and type of perturbation~\citep{elenius2014}. While our choice of
perturbation initially has a grid dependence, the shortest wavelength introduced
by the grid is significantly shorter than the first unstable mode from the linear
stability analysis and therefore decay quickly until the onset of the nonlinear
viscous fingering. If the onset time of the unstable viscous fingering is of
particular interest, more care is needed.

\subsection{Dimensionless equations}\label{sec:scaling}
The characteristic length scale for our problem is given by the width of the domain $H$ as shown in Figure~\ref{fig:fractured_domain}. The characteristic velocity is chosen as the matrix injection flux
$U_m$, while the characteristic viscosity is $\mu_d$. By following classical
results from the study of viscous fingering (see, e.g., \cite{tan1986stability}), the P{\'
  e}clet number is defined by $\Pen = \frac{U_mH}{D}$, the characteristic time by
$T = \frac{\phi_m H}{U_m}$, and the characteristic pressure by
$P = \frac{\mu_d U_mH}{k_m}$. The dimensionless variables are denoted by a hat
$\hat\cdot$, that by a rescaling of Equation~\eqref{eq:matrix_equation} gives the
dimensionless equations defined in the rock matrix:
\begin{equation}\label{eq:matrix_equation_dimless}
  \begin{aligned}
    \hat{\vec u}_m = -\frac{1}{\hat\mu(c)}\nabla \hat p_m, \quad \nabla \cdot \hat{\vec u}_m = 0\\
    \frac{\partial c_m}{\partial \hat t} + \nabla \cdot c_m\hat{\vec u}_m -
    \frac{1}{\Pen}\Delta c_m = 0
  \end{aligned}
  \qquad \text{in}\ \hat\Omega_m.
\end{equation}

Similarly, in the fractured domain the flux is rescaled by the characteristic
fracture flux $HU_m$. We will assume that the fractures are
filled with a material that has the same porosity and diffusivity as the rock
matrix, i.e., $\phi_f = \phi_m$, $D_f = D_m$. This assumption on the diffusion in
the fractures is not valid in general, however, we are interested in the case of
highly permeable fractures where advection dominates over diffusion in the
fractures and the fracture diffusion plays a minor role. This gives the
dimensionless equations:
{\par\nobreak\noindent}
\begin{align}\label{eq:fracture_equation_dimless}
  \begin{aligned}
    \hat{\vec u}_f = -\frac{\mathcal{KA}}{\hat\mu(c_f)}\nabla_{||} \hat p_f, \quad \mathcal
    \nabla_{||} \cdot \hat{\vec u}_f
    -[\![\hat{\lambda}]\!]= 0 \\
    \frac{\mathcal A\partial c_f}{\partial \hat t} + \nabla_{||} \cdot c_f\hat{\vec u}_f - \frac{\mathcal A}{\Pen}\Delta_{||} c_f
    - [\![\hat\lambda_c]\!] = 0
  \end{aligned}
  \qquad &\text{in}\ \hat\Omega_f,
\end{align}
where we have defined the dimensionless numbers
\begin{equation}\label{eq:KW}
  \quad \mathcal K = \frac{k_f}{k_m},%
  \quad \mathcal A = \frac{a}{H}.
\end{equation}
The permeability ratio, $\mathcal K$, defines the ratio of the permeability in the
fractures and rock matrix, while the second dimensionless number, $\mathcal A$, is the
dimensionless aperture which represents the ratio of the width of the fractures and
the characteristic width of the domain that is related to the P{\' e}clet number.

The dimensionless coupling equations are obtained by rescaling
Equations~\eqref{eq:advective_coupling} and~\eqref{eq:coupling_flux}:
\begin{equation} \label{eq:coupling_flux_dimless}
  \begin{aligned}
    \hat{\lambda}^\pm %
    = -\frac{\mathcal K}{\mathcal A\hat\mu(c^\pm)}%
    \frac{\hat p_f - \text{tr}(\hat p_m)}{1 / 2},\quad%
    \hat \lambda_c^\pm = c^\pm\hat {\lambda}^\pm - \frac{1}{\mathcal A\Pen} \frac{c_f - \text{tr}(c_m)}{1 / 2}
  \end{aligned}
  \quad \text{on}\ \hat{\Gamma}^\pm.
\end{equation}

To summarize, the dimensionless variables are
{\par\nobreak\noindent}
\begin{equation*}
  \begin{aligned}
    \hat{\vec x} &= \frac{1}{H}\vec x,%
    \quad &\hat t &= \frac{U_m}{\phi_m H} t,%
    \quad &\hat\mu &= \frac{1}{\mu_d}\mu,%
    \quad &\hat\lambda &= \frac{1}{U_m}\lambda\\
    \hat{\vec u}_m &= \frac{1}{U_m}\vec u_m,%
    \quad &\hat{\vec u}_f &= \frac{1}{HU_m}\vec u_f,%
    \quad & \hat p_m & = \frac{k_m}{\mu_dU_mH}p_m,%
    \quad &\hat p_f &= \frac{k_m}{\mu_dU_mH}p_f.%
  \end{aligned}
\end{equation*}
From the dimensionless
Equations~\eqref{eq:matrix_equation_dimless}-\eqref{eq:coupling_flux_dimless}, we
obtain that the dimensionless numbers governing the behavior of this system:
\begin{equation}\label{eq:dimless_numbers}
  \Pen = \frac{U_mH}{D},\quad
  R = \ln\left(\frac{\mu_d}{\mu_u}\right),\quad
  \mathcal K = \frac{k_f}{k_m},\quad
  \mathcal A = \frac{a}{H},\quad
  N = \frac{H}{W}.
\end{equation}
In addition, we have here introduced the dimensionless number, $N$, that represents 
the fracture density in a vertical slice of the domain. While the fracture density, 
$N$, to some extent describe the fracture network geometry, it does not describe it 
fully. The details of the fracture geometry is given in  Section~\ref{sec:hom_and_layered}.

\subsection{Quantitative measures}\label{sec:quantitative_measures}
As a quantitative measure of the global solution, we use the mixing length $h(\hat t)$
and the number of fingers in the domain $n(\hat t)$. The mixing length is defined as the
length where the two fluids have spread:
\begin{equation}
  \label{eq:finger_length}
  h(\hat t) = \max(\hat x|_{c(\hat x, \hat y, \hat t)>0.01 }) - \min(\hat x|_{c(\hat x, \hat
    y, \hat t)<0.99}).
\end{equation}
The second quantitative measure, the number of fingers, is calculated as
{\par\nobreak\noindent}
\begin{equation*}
  n(\hat t) = \frac{1}{h(\hat t)}\int_{\min(\hat x_{\hat x, \hat
      y, \hat t)<0.99})}^{\max(\hat x|_{c(\hat x, \hat y, \hat t)>0.01 })} \eta(x)\d x,
\end{equation*}
where $\eta(x)$ is the number of local maxima in a vertical slice.

\section{Numerical methods}\label{sec:numerical_method}
The numerical schemes used for studying fingering phenomena in porous media are
typically higher order schemes~\citep{riaz2006,nijjer2019stable,wit1997viscous,sajjadi2013scaling}. While these methods are superior on regular domains, they are challenging to
employ in domains with complex fracture geometries, which is one of the main concerns in this paper. Further, the higher order schemes usually require a smooth
permeability field, while the contrast between the fractures and rock matrix permeability represents a discontinuity. The final case presented in this paper necessitates the use of numerical schemes that can handle complex unstructured fracture network geometries, hence, in this paper we use a finite-volume scheme that,
in addition to handling discontinuities in the permeability field, is easy to implement for the lower-dimensional fractures. The reduction of the dimension
of the fractures allows us to simulate on layers that are orders of magnitude thinner
than the domain width without needing an extensive grid refinement around the fractures, which makes it feasible to run long time simulations. In this section, the discretization is presented, as well as details about the mesh adaptivity and the implementation.

Each subdomain, $\hat \Omega_m$ and $\hat \Omega_f$, is partitioned into a set of
non-overlapping cells that defines one mesh of the 2D domain and one mesh of the
1D domain. The 2D mesh is a logical Cartesian mesh, and we use a 1D mesh with cells
that conform to the faces of the 2D mesh.

\subsection{Spatial and temporal discretization}
The system of Equations~\eqref{eq:matrix_equation_dimless}-\eqref{eq:coupling_flux_dimless}
is approximated using a finite volume discretization. The flux from the rock matrix to
the fracture domain, defined in
Equation~\eqref{eq:coupling_flux_dimless}, is included as a Neumann boundary in the rock
matrix and a source term in the fractures. For details on the mixed dimensional
coupling we refer the interested reader to~\citep{keilegavlen2021}. As the discretizations
of the rock matrix domain and the fracture domain are equivalent, we drop the domain subscript
$m$ and $f$ when we describe the discretization in the following subsections.

\subsubsection*{Pressure equation}
To discretize the pressure equation, Darcy's law is substituted into the mass concervation equation
and integrated over a cell $L$. By applying the divergence theorem we obtain:
\begin{equation}
\int_{\partial L} \frac{k}{\hat\mu(c)}\nabla \hat p \cdot \vec \nu \d S - \int_L Q\d V = 0,
\end{equation}
where the term $Q$ is a source term and $\vec\nu$ is the outwards pointing normal vector of $\partial L$. Note also that the permeability in the rock matrix for the dimensionless equation is $1$ while it is $\mathcal{KA}$ in the fractures. 
Further, the source term is zero in the rock matrix, and $Q=[\![\hat \lambda]\!]$ in the fractures. The integral of the source term over the cell is approximated by multiplying
the cell-centered value of the source by the cell volume. To discretize the flux term, let
$\sigma$ be the face between cell $L$ and cell $R$ (referred to as left and right cell). The fluid flux across the face is approximated as
\begin{equation}\label{eq:viscous_flux_disc}
      \int_\sigma \frac{k}{\hat \mu(c)}\nabla\hat p \cdot \vec \nu \d S\approx F_{LR} =\frac{T_{\sigma}}{\hat\mu(c_{\sigma})} (\hat p_L-\hat p_R),
\end{equation}
where $T_\sigma$ is the transmissibility and $\hat p_i$ is the cell center pressure of cell $i\in \{L, R\}$. All numerical simulations in this paper are run on a Cartesian grid, and we use the Two-Point Flux Approximation (TPFA) to calculate the transmissibility. In TPFA, the transmissibility is calculated as the harmonic average of the half-transmissibilities: 
{\par\nobreak\noindent}
\begin{equation*}
    T_{\sigma} = \frac{T_{\sigma L}T_{\sigma R}}{T_{\sigma L}+T_{\sigma R}},\quad T_{\sigma i} = \frac{|\sigma|k\vec d_{\sigma i} \cdot \vec\nu_{\sigma i}}{\vec d_{\sigma i}\cdot \vec d_{\sigma i}}, \quad i\in\{L, R\},
\end{equation*}
where $|\sigma|$ is the area of the face, $\vec d_{\sigma i}$ is the vector pointing from the cell-center of cell $i$ to the face center of face $\sigma$, and $\vec \nu_{\sigma i}$ is the normal vector of face $\sigma$ pointing outwards of cell $i$. The concentration on the face, $c_{\sigma}$, is taken as the upwind value of the concentration:
\begin{equation}\label{eq:upwind}
    c_{\sigma} = \text{up}(c, \hat p_L, \hat p_R) =
    \begin{cases}
    c_L, &\text{if}\ \hat p_L-\hat p_R > 0\\
    c_R, &\text{if}\ \hat p_L-\hat p_R \le 0,
    \end{cases}
\end{equation}
for the cell center concentrations $c_L$, and $c_R$.

\subsubsection*{Transport equation}
A similar approach is used for the transport equation where we first integrate over a cell $L$:
\begin{equation}\label{eq:transport_integrated}
  \int_L \frac{\partial \phi c}{\partial \hat t}\d V - \int_{\partial L} (c\hat{\vec u} -
  D\nabla c)\cdot \vec\nu \d S - \int_L Q_c\d V = 0,
\end{equation}
where $Q_c = 0$, $\phi=1$ in the rock matrix and $Q_c = [\![\hat \lambda_c]\!]$, $\phi =\mathcal A$ in the fractures (for the dimensionless equation). The diffusive flux
$\int_{\partial L}D\nabla c\cdot \vec\nu \d S$ is approximated by the TPFA scheme, equivalent to Equation~\eqref{eq:viscous_flux_disc} ($D=1/\Pen$ in $\hat \Omega_m$ and $D=\mathcal A/\Pen$ in $\hat \Omega_f$). The advective flux accross the face $\sigma$ is approximated by
\begin{equation}
    \int_{\sigma} c\hat{\vec u}\cdot \vec\nu\d S \approx c_{\sigma} F_{LR},
\end{equation}
where $F_{LR}$ is the flux defined by Equation~\eqref{eq:viscous_flux_disc} and $c_{\sigma}$ is the upwind value as defined by Equation~\eqref{eq:upwind}. For the time derivative of Equation~\eqref{eq:transport_integrated} the second-order Crack-Nicholson scheme is used.

\subsubsection*{Coupling equations}
To discretize the coupling equations~\eqref{eq:coupling_flux_dimless}
a discrete trace operator in the rock matrix must be defined. The TPFA scheme allows for the reconstruction of the pressure/concentration on the faces of the cells from the half-transmissibilities, and we employ this reconstruction as the trace operator. To evaluate the viscosity on the interface we use the upstream value of the concentration in the rock matrix and in the fracture.

\subsubsection*{Linearization and time stepping}
The coupling between the transport equation, the pressure equation and the concentration dependent viscosity results in a discrete non-linear system of equations. The non-linear system is
linearized by Newton's method, and the Jacobian is obtained by forward-mode Automatic Differentiation~\citep{neidinger2010}.

The initial time-step size is set to
$\Delta \hat t_0 = 10^{-5}$, and the time-step size is incremented by a constant
factor after each time step until it reaches the maximum time-step size,
$\Delta \hat t_{\max} = 10^{-2}$. The increment factor is $1.2$ if the Newton
scheme converges in less than 3 iterations, $1.1$ if the Newton scheme converges
in less than 7 iterations and there is no time-step size increment if the Newton
iteration takes more than or equal 7 iterations. If the Newton iteration fails the time step is halved and the iteration restarted.

\subsection{Mesh adaptivity}
The initial condition contains a sharp
concentration front, and linear stability analysis shows that many small fingers
(the number related to the most stable wave number) will appear at the onset of
the viscous fingering~\citep{tan1986stability}. Thus, a very fine mesh is needed
to resolve the initiation of these fingers accurately. In the simulations in this paper we use an initial mesh with a cell length $\sim 0.002$. At the shutdown of the viscous fingering, the finger length is typically of order 100, and even higher for the cases with high permeability contrast. If the initial mesh size is kept for the whole simulations this would result on the order of $10^7$ cells. Thus, in order to simulate for a long time scale, $\hat t > 10$, adaptivity of the mesh
is needed to achieve acceptable run-times. We employ three adaptive strategies to make the runtimes feasible. The first adaptive strategy is to extend the simulation domain when the viscous
fingers grows to close to the domain boundary. In the simulations, the domain is
extended if the viscous fingers are closer to the domain boundary than 20 \% of the current domain
length. The second adaptive strategy employed is to coarsen the mesh size when the
fine mesh is no longer needed; as the concentration front diffuses, the
displacement front smoothens out which allows for a coarser mesh. In the numerical
simulations the mesh is coarsened when the extension of the grid causes the number
of cells to exceed 600 000. When the domain is coarsened we also double the
maximum allowed time step size in the temporal discretization. The final adaptive strategy is to shift the domain downstream if the distance from the upstream boundary to the concentration front ($\min(\hat x|_{c(\hat x,\hat y,\hat t)<0.99))}$) is longer than 30 \% of the domain length. We have compared
the solutions of several simulations using the adaptive strategies to the solutions
on a fine mesh and not found any notable discrepancies between the solutions that have a significant impact on either the qualitative structure of the fingers, nor the measurables defined in Section~\ref{sec:quantitative_measures}

\subsection{Implementation and software}
The implementation is done in the open-source software
PorePy~\citep{keilegavlen2021}, and we have validated our numerical method by
comparing the initial fingering in a domain without fractures to both linear stability
analysis and to the evolution of the viscous and late time regimes described in the
numerical simulations by~\citet{nijjer_2018}. The mixed-dimensional
coupling between the fractures and rock matrix has been applied in PorePy for both
flow and transport in several
papers~\citep{nordbotten2019unified,stefansson2018,fumagalli2019}. To solve the resulting linear system the direct solver UMFPACK is used~\citep{davis2004}, and Paraview~\citep{ayachit2015paraview} has been applied for
postprocessing and visulalization of the results. To facilitate reproducibility, the
run-scripts used in the simulations reported below are available at GitHub\footnote{https://github.com/rbe051/ViscFrac.git} and archived on Zenondo~\citep{berge2023}. In addition, the results of the simulations are also archived on Zenondo~\citep{berge2023b}

\section{Numerical simulations}\label{sec:hom_and_layered}
\begin{figure}
  \centering \def\figWidth{0.23\textwidth}
  \begin{subfigure}[b]{\figWidth}
    \centering \def\svgwidth{\textwidth}
\begingroup%
  \makeatletter%
  \providecommand\color[2][]{%
    \errmessage{(Inkscape) Color is used for the text in Inkscape, but the package 'color.sty' is not loaded}%
    \renewcommand\color[2][]{}%
  }%
  \providecommand\transparent[1]{%
    \errmessage{(Inkscape) Transparency is used (non-zero) for the text in Inkscape, but the package 'transparent.sty' is not loaded}%
    \renewcommand\transparent[1]{}%
  }%
  \providecommand\rotatebox[2]{#2}%
  \newcommand*\fsize{\dimexpr\f@size pt\relax}%
  \newcommand*\lineheight[1]{\fontsize{\fsize}{#1\fsize}\selectfont}%
  \ifx\svgwidth\undefined%
    \setlength{\unitlength}{302.83463125bp}%
    \ifx\svgscale\undefined%
      \relax%
    \else%
      \setlength{\unitlength}{\unitlength * \real{\svgscale}}%
    \fi%
  \else%
    \setlength{\unitlength}{\svgwidth}%
  \fi%
  \global\let\svgwidth\undefined%
  \global\let\svgscale\undefined%
  \makeatother%
  \begin{picture}(1,1)%
    \lineheight{1}%
    \setlength\tabcolsep{0pt}%
    \put(0,0){\includegraphics[width=\unitlength,page=1]{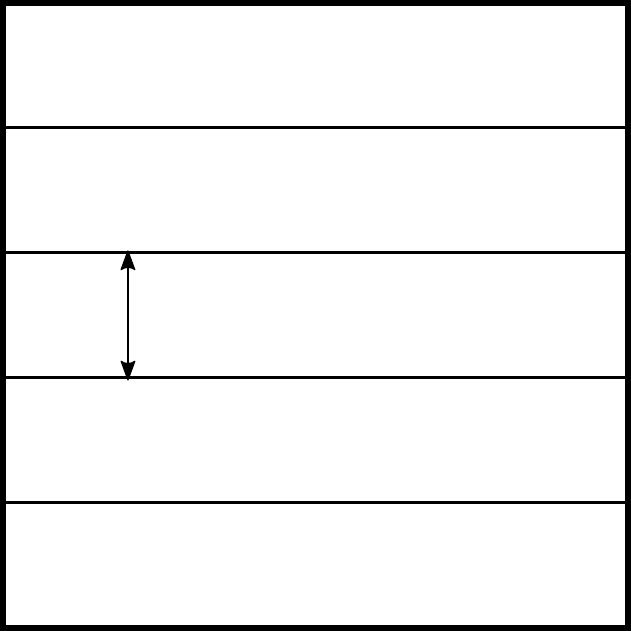}}%
    \put(0.22262088,0.47976477){\color[rgb]{0,0,0}\makebox(0,0)[lt]{\lineheight{1.25}\smash{\begin{tabular}[t]{l}$W$\end{tabular}}}}%
  \end{picture}%
\endgroup%

    \caption{Parallel}
  \end{subfigure}
  \hfill
  \begin{subfigure}[b]{\figWidth}
    \centering \def\svgwidth{\textwidth}
\begingroup%
  \makeatletter%
  \providecommand\color[2][]{%
    \errmessage{(Inkscape) Color is used for the text in Inkscape, but the package 'color.sty' is not loaded}%
    \renewcommand\color[2][]{}%
  }%
  \providecommand\transparent[1]{%
    \errmessage{(Inkscape) Transparency is used (non-zero) for the text in Inkscape, but the package 'transparent.sty' is not loaded}%
    \renewcommand\transparent[1]{}%
  }%
  \providecommand\rotatebox[2]{#2}%
  \newcommand*\fsize{\dimexpr\f@size pt\relax}%
  \newcommand*\lineheight[1]{\fontsize{\fsize}{#1\fsize}\selectfont}%
  \ifx\svgwidth\undefined%
    \setlength{\unitlength}{302.83463125bp}%
    \ifx\svgscale\undefined%
      \relax%
    \else%
      \setlength{\unitlength}{\unitlength * \real{\svgscale}}%
    \fi%
  \else%
    \setlength{\unitlength}{\svgwidth}%
  \fi%
  \global\let\svgwidth\undefined%
  \global\let\svgscale\undefined%
  \makeatother%
  \begin{picture}(1,1)%
    \lineheight{1}%
    \setlength\tabcolsep{0pt}%
    \put(0,0){\includegraphics[width=\unitlength,page=1]{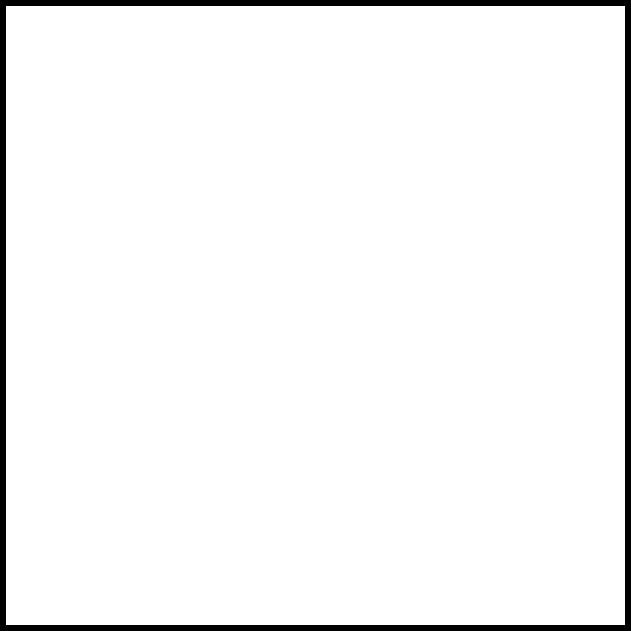}}%
    \put(0.0957837,0.37665368){\color[rgb]{0,0,0}\makebox(0,0)[lt]{\lineheight{1.25}\smash{\begin{tabular}[t]{l}$W$\end{tabular}}}}%
    \put(0,0){\includegraphics[width=\unitlength,page=2]{fracture_networks_brick.pdf}}%
  \end{picture}%
\endgroup%

    \caption{Brick\label{fig:fracture_networks_brick}}
  \end{subfigure}
  \hfill
  \begin{subfigure}[b]{\figWidth}
    \centering \def\svgwidth{\textwidth}
\begingroup%
  \makeatletter%
  \providecommand\color[2][]{%
    \errmessage{(Inkscape) Color is used for the text in Inkscape, but the package 'color.sty' is not loaded}%
    \renewcommand\color[2][]{}%
  }%
  \providecommand\transparent[1]{%
    \errmessage{(Inkscape) Transparency is used (non-zero) for the text in Inkscape, but the package 'transparent.sty' is not loaded}%
    \renewcommand\transparent[1]{}%
  }%
  \providecommand\rotatebox[2]{#2}%
  \newcommand*\fsize{\dimexpr\f@size pt\relax}%
  \newcommand*\lineheight[1]{\fontsize{\fsize}{#1\fsize}\selectfont}%
  \ifx\svgwidth\undefined%
    \setlength{\unitlength}{302.83463125bp}%
    \ifx\svgscale\undefined%
      \relax%
    \else%
      \setlength{\unitlength}{\unitlength * \real{\svgscale}}%
    \fi%
  \else%
    \setlength{\unitlength}{\svgwidth}%
  \fi%
  \global\let\svgwidth\undefined%
  \global\let\svgscale\undefined%
  \makeatother%
  \begin{picture}(1,1)%
    \lineheight{1}%
    \setlength\tabcolsep{0pt}%
    \put(0,0){\includegraphics[width=\unitlength,page=1]{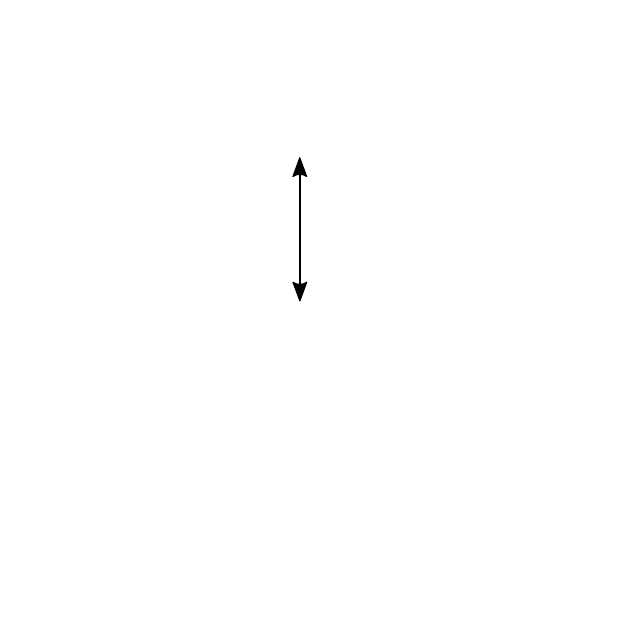}}%
    \put(0.495647,0.62263293){\color[rgb]{0,0,0}\makebox(0,0)[lt]{\lineheight{1.25}\smash{\begin{tabular}[t]{l}$W$\end{tabular}}}}%
    \put(0,0){\includegraphics[width=\unitlength,page=2]{fracture_networks_random.pdf}}%
  \end{picture}%
\endgroup%

    \caption{Random\label{fig:fracture_networks_random}}
  \end{subfigure}

  \caption{Schematics of the fracture networks studied in the numerical simulations. The characteristic width $W$ is indicated by the double arrow. The principle direction of flow is from left to right.}
  \label{fig:fracture_networks}
\end{figure}

In the following subsections we study the three different characteristic fracture network geometries illustrated in Figure~\ref{fig:fracture_networks}. The fracture networks are chosen to illustrate a wide range of different flow dynamics and interactions
between the fracture and rock matrix. The following test case with only fractures parallel to the flow directions is equivalent to a layered porous media where the width of the high-permeability layer (the fracture) is several orders of magnitude smaller than the width of the low-permability layer (the matrix). Compared to the parallel fractures, the brick network and random network have a
complex geometry with intersecting fractures, fractures of finite size and fractures not aligned with the flow direction.

\subsection{Parallel fractures}\label{sec:parallel}
To understand the effect fractures have on viscous fingering, we first investigate the
geometrically simple case of fractures parallel to the flow direction. The key question
we study is how the dimensionless numbers $\mathcal K$, $\mathcal A$ and $N$ in Equation~\eqref{eq:dimless_numbers},
which are introduced by the addition of fractures, affect the viscous fingering compared to a homogeneous porous medium. 

All simulations in this subsection are run in a moving frame of
reference by a change of variables:
{\par\nobreak\noindent}
\begin{equation*}
  \tilde x = \hat x - \hat t,\  \tilde{\vec u} = \hat{\vec u} - \vec i,
\end{equation*}
where $\vec i$ is the unit basis vector in the $\hat x$-direction. The initial
dimensionless size of the computational domain is set to $L / H = 2$.

\subsubsection{Influence of permeability ratio and dimensionless aperture} \label{sec:paralell_KA}%
We investigate how the viscous fingering is affected by the permeability ratio and
dimensionless aperture by fixing the P{\' e}clet number to $1000$ and the log viscosity
ratio to $R=2$. The domain contains a single fracture located in the middle of the
domain. The permeability ratio is varied between $\mathcal K=1.1$ and $\mathcal K=10^3$, and the
dimensionless aperture is varied between $\mathcal A = 10^{-2}$ and
$\mathcal A = 10^{-4}$.

\begin{figure}
  \centering
  \begin{minipage}[c]{.98\textwidth}
  \def\figWidth{0.5\textwidth}
  \begin{subfigure}[b]{\figWidth}
      \includegraphics[width=\textwidth]{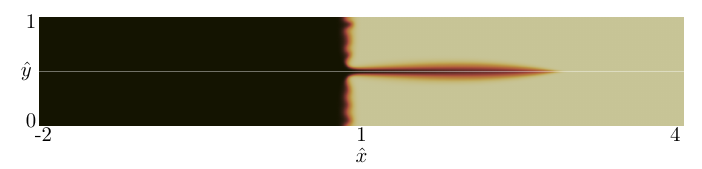}
    \caption{ $\hat t = 1$}
  \end{subfigure}%
  \begin{subfigure}[b]{\figWidth}
      \includegraphics[width=\textwidth]{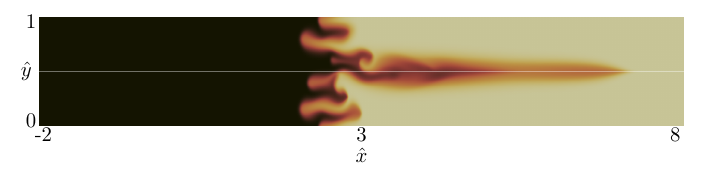}
    \caption{ $\hat t = 3$}
  \end{subfigure}
  \begin{subfigure}[b]{\figWidth}
    \includegraphics[width=\textwidth]{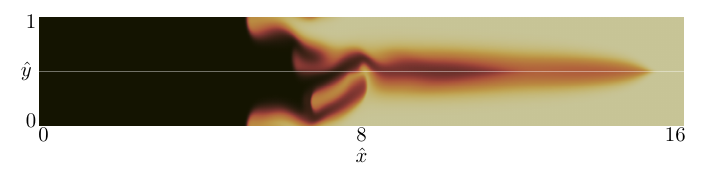}
    \caption{$\hat t = 8$}
  \end{subfigure}%
  \begin{subfigure}[b]{\figWidth}
    \includegraphics[width=\textwidth]{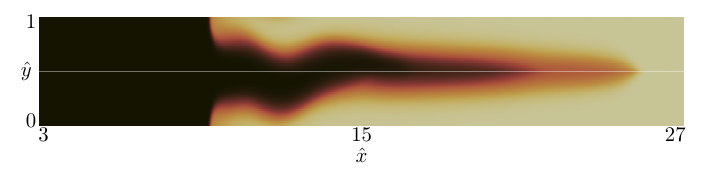}
    \caption{$\hat t = 15$}
  \end{subfigure}%
  \end{minipage}%
  \begin{minipage}[c]{.02\textwidth}
  \begin{subfigure}[b]{\textwidth}
  \includegraphics[width=\textwidth]{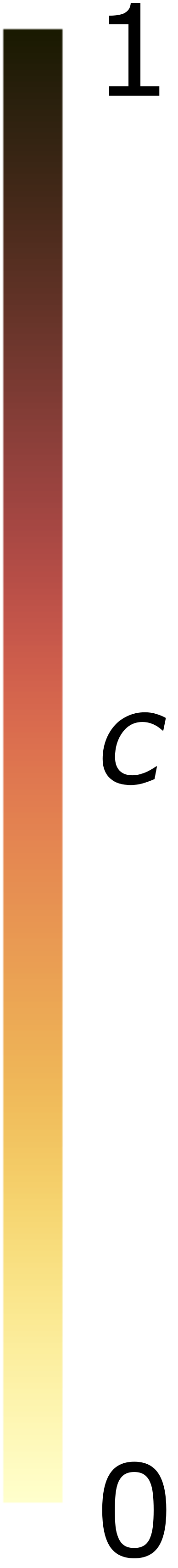}
  \end{subfigure}%
  \end{minipage}%
  \caption{Evolution of the concentration field for the parameters
    $\{\Pen, R, \mathcal K, \mathcal A, N\} = \{1000, 2, 10, 10^{-3}, 1\}$. The white line indicates the position of the fracture. Note the increasing aspect ratio in the figures.}
  \label{fig:hfrac_evolution}
\end{figure}

Initially, the concentration is transversely homogeneous with a sharp concentration front at $x=0$. The diffusion across the concentration front dominates and the transport equation in the fracture and the rock matrix decouples and reduces to a pure diffisive transport. Thus, at early times the mixing length that scales as $h\sim\sqrt{\hat t/\Pen}$. The difference in permeability between the fracture and the rock matrix causes an advective growth of the mixing length, $h=O\left((\mathcal K-1)\hat t\right)$, that outgrows the diffusive initial regime. In all simulations an initial finger develops around the fracture that outgrows any fingers caused solely by the viscous instabilities. Figure~\ref{fig:hfrac_evolution} shows the continued evolution of the concentration front that is typical for the cases where the permeability ratio is less than 10. In these cases, we observe that the viscous fingers exhibit growth patterns similar to those observed in homogeneous domains within the rock matrix, away from the fractures. Following the initiation of viscous fingers in the rock matrix, the viscous fingers in the rock matrix compete with the finger aligned with the fracture before undergoing coarsening and merging processes, eventually leaving only a single dominant finger in the domain.

\begin{figure}
  \centering 
  \includegraphics[width=1\textwidth]{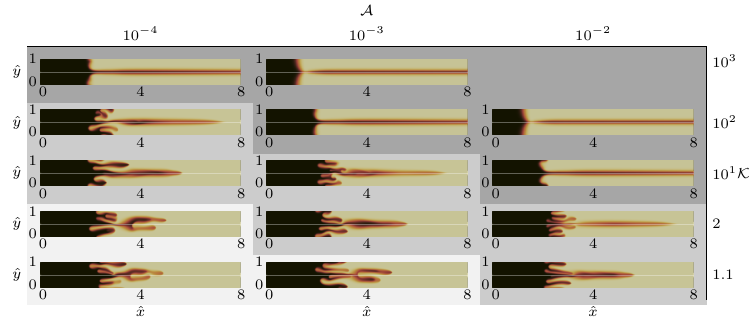}
  \includegraphics[width=0.2\textwidth]{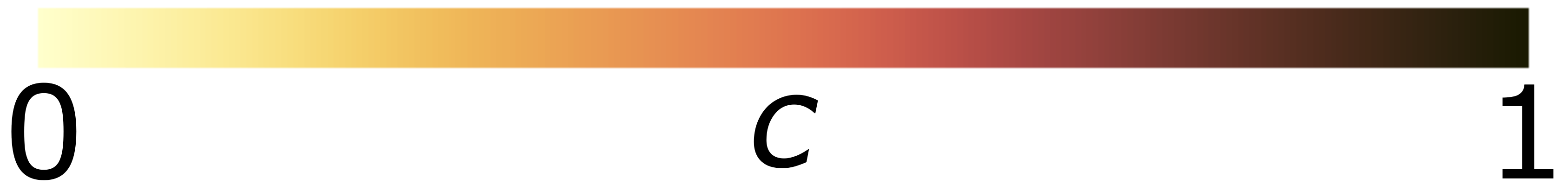}
  \caption{Domain with a horizontal fracture shown as a white line. The figure
    shows the concentration for $\hat t=3$, for different permeability ratios and
    dimensionless apertures. The other parameters are fixed to
    $(R, \Pen , N) = (2, 1000, 1)$. The different background shades of gray indicate different qualitative regimes. Note that the simulation domains are much
    larger than shown in the figures. \label{fig:parallel_KvsA_snap}}
\end{figure}

For all parameter ranges investigated in this study, a common observation is the initial growth of a finger aligned with the fracture and the final regime with a single propagating finger. However, there are variations in the behavior of the simulations during intermediate times. Particularly, for cases with the largest permeability ratio we observe that the viscous fingers in the rock matrix do not exhibit significant growth. Instead, the flow transitions directly from a diffusive regime to the dominance of a single propagating finger. This is illustrated in Figure~\ref{fig:parallel_KvsA_snap} that shows snapshots of the concentration profiles of all simulations at time $\hat t=3$. We have divided the flow into three qualitative regimes, indicated by the different background shades of gray, which we will elaborate on in the following paragraphs.

For small permeability ratio and small dimensionless aperture the viscous fingering dominates over the fracture flow, indicated by the lightest shade (bottom left). In these simulations the viscous fingering in the rock matrix has displaced the finger aligned with the fracture. For the largest permeability ratios and highest dimensionless apertures the finger aligned with the fracture suppresses the growth of the viscous fingers, and we observe no viscous fingering in the rock matrix except the one aligned with the fracture, indicated by the darkest shade (upper right). In the intermediate parameter regime the dominant finger aligned with the fracture is competing with the viscous fingering in the rock matrix (indicated by the intermediate shade). The fingers that develops in the rock matrix transition the flow to a regime where the viscous fingers dominates over the fracture.

In general, the qualitative classification into these three regimes is not stable through the whole simulation, but may transition from one regime to another. For the parameters we have studied we have only observed the fluid flow evolve from a more fracture dominated flow to a more viscous fingering dominated flow as a function of time.

\begin{figure}[h]
  \centering \def\figWidth{1\textwidth}
  \begin{subfigure}[b]{\figWidth}
    \centering \includegraphics[width=0.75\textwidth]{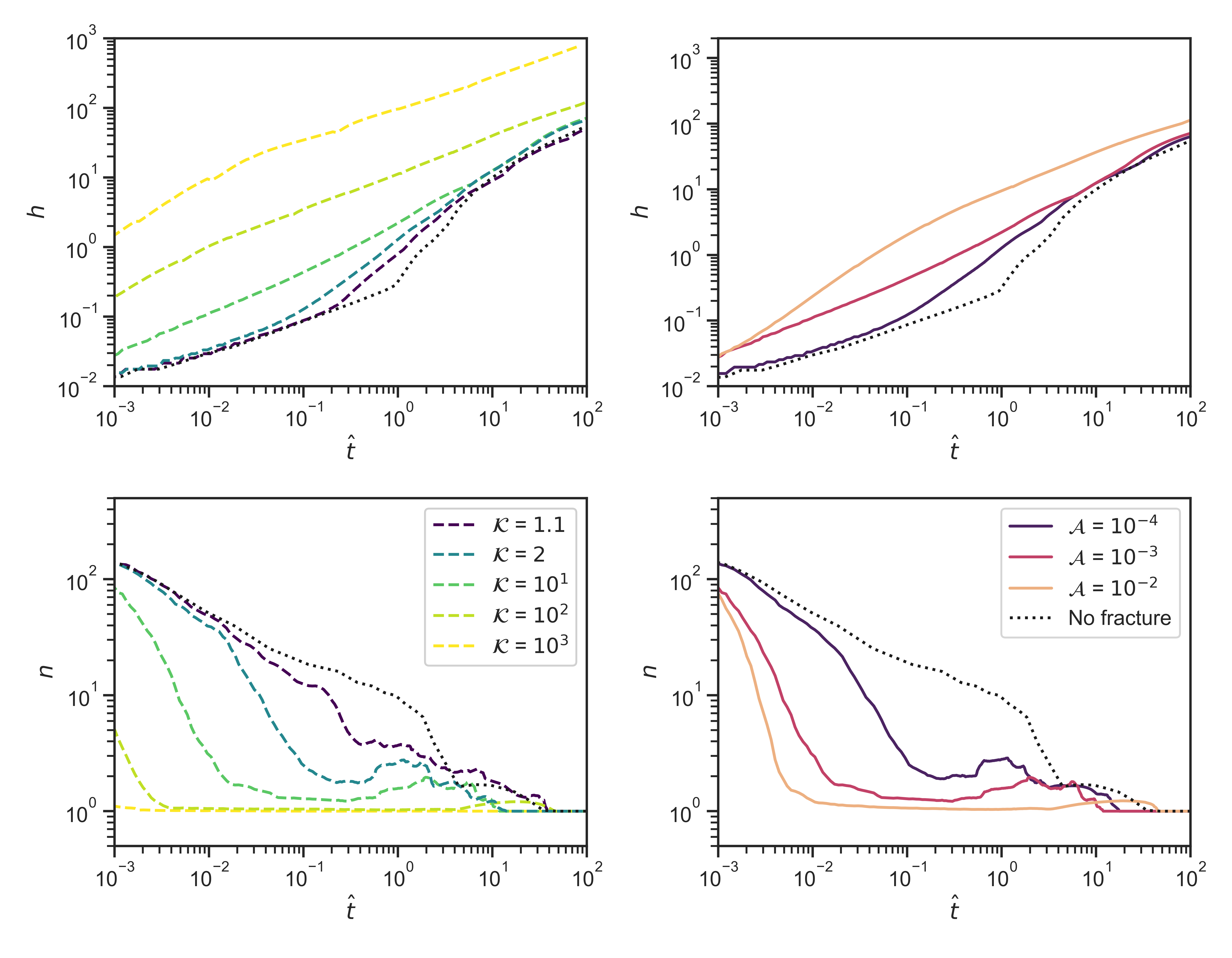}
  \end{subfigure}
  \caption{Evolution of the quantitative measures for a domain with a single
    fracture parallel to the flow direction. The P{\' e}clet number, log viscosity ratio
    and fracture density are $\{Pe, R, N\} = \{1000, 2, 1\}$. The dotted black
    line represents measures from a simulation without any fractures. The left column shows the mixing length and number of
    fingers for different permeability ratios $\mathcal K$ and fixed dimensionless
    aperture $\mathcal A=10^{-3}$. The right column shows the mixing length and number of fingers for
    fixed permeability ratio $\mathcal K=10$ and different dimensionless aperture $\mathcal A$.
  }
  \label{fig:hfrac_KvsA}
\end{figure}

The effect of the fracture can also be seen in the quantitative measures plotted in Figure~\ref{fig:hfrac_KvsA}, which plot the mixing length for varying permeability ratios and dimensionless aperture. Even for the smallest permeability ratio of the simulations, $\mathcal K=1.1$, the number of fingers is reduced by almost an order of magnitude compared to the case without fractures at $\hat t\approx 1$. As the viscous fingering develops, we can also observe the transition from a fracture dominated flow to a viscous dominated flow in the quantitative measures as an increase in the number of fingers. 

\begin{figure}
  \centering%
  \def\figWidth{1\textwidth}
  \begin{subfigure}[T]{0.8\figWidth}
    \includegraphics[width=1\textwidth]{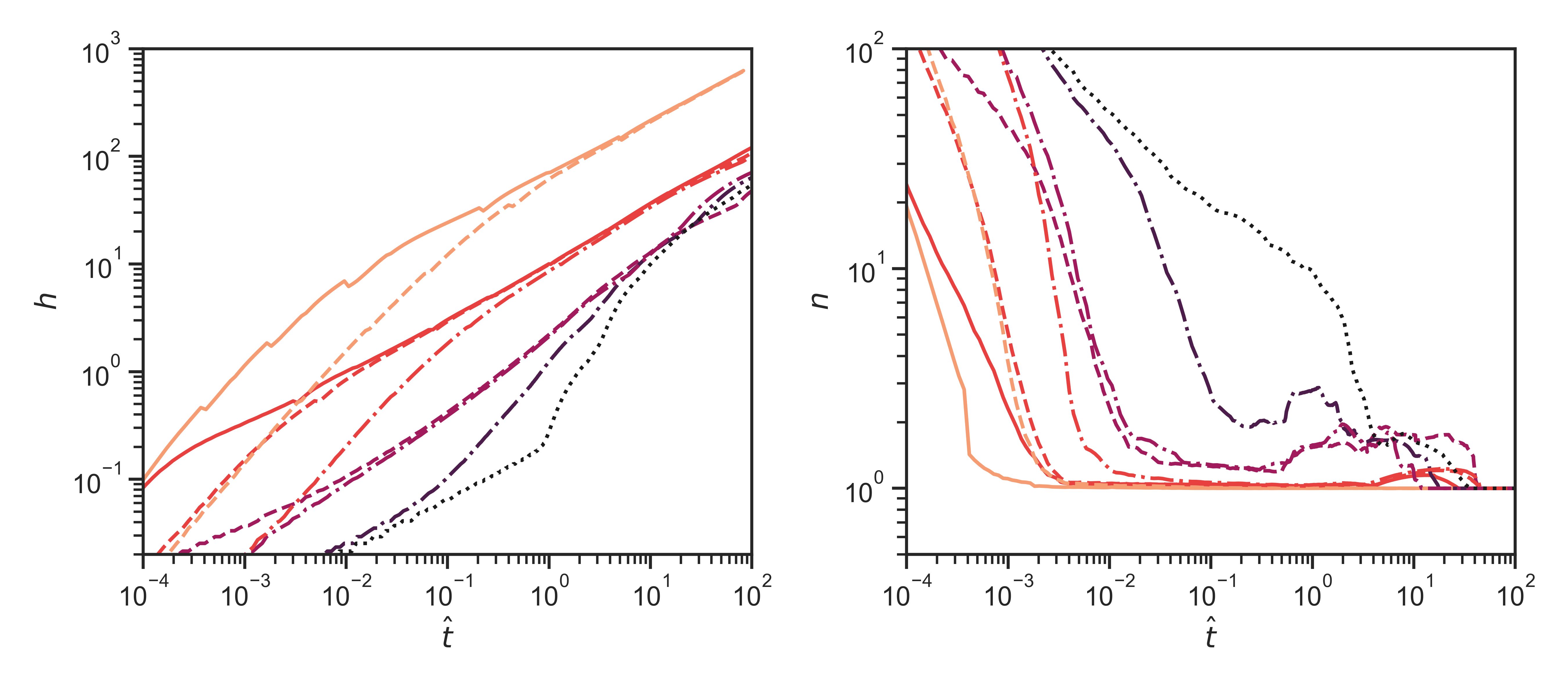}
  \end{subfigure}
  \hspace{-1em}
  \begin{subfigure}[T]{\legendWidth}
    \vspace{1em}
    \includegraphics[width=1\textwidth]{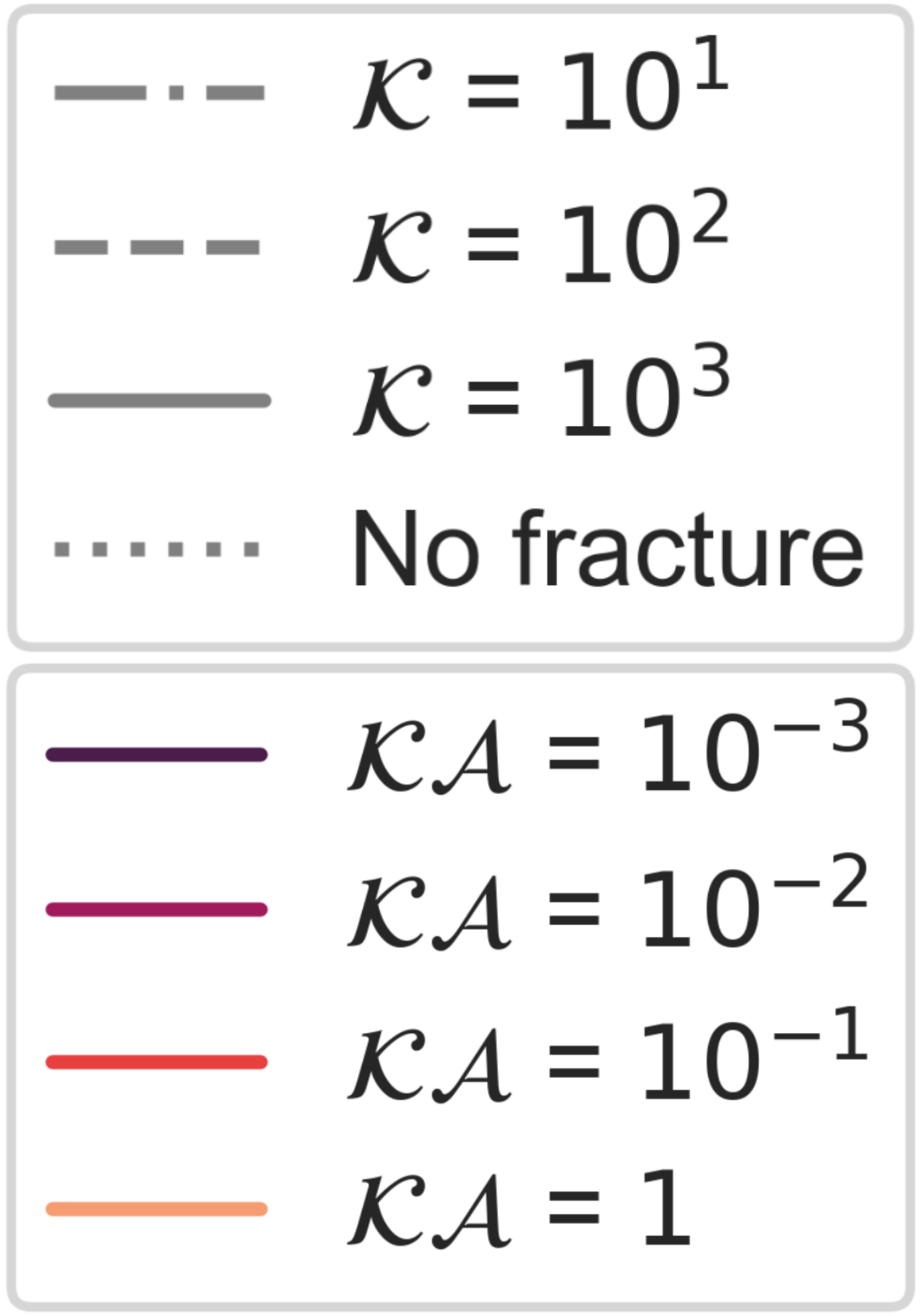}
    \end{subfigure}
  \caption{Evolution of the concentration field for a domain with a single
    fracture parallel to the flow direction. The P{\' e}clet number, log viscosity ratio
    and fracture density are $\{Pe, R, N\} = \{1000, 2, 1\}$. The dotted black
    line is a simulation without any fractures. The colors show the volumetric
    flux ratio $\mathcal K\mathcal A$, and the line style the value of $\mathcal
    K$. Left figure shows the mixing length and right figure shows the number of fingers. The small irregularities for the case $\mathcal K=10^3$, $\mathcal{KA}=1$ is due to the coarsening of the mesh, however, we have no reason to believe that this affect the general conclusions.
  }
  \label{fig:hfrac_F}
\end{figure}

As inferred from Figures 4 and 5, the quantity $\mathcal{KA}$ both scales the fluid flux in the fracture, and also to first order describes the qualitative structure of the solution. The dimensionless number $\mathcal{KA}$ represents the ratio of the volumetric flux in the fractures to the volumetric flux in the rock matrix, and we expect it to be important when describing the behaviour of the flow. Figure~\ref{fig:hfrac_F} plots the mixing length and number of fingers for the volumetric flux ratio varying from $\mathcal K\mathcal A=10^{-3}$ to $\mathcal K\mathcal A = 1$.
For each value of the volumetric flux ratio, the permeability and aperture is varied three orders of magnitude. Due to the initial conditions, the fluid flux between the rock matrix and the fracture is initially zero and the equations in the rock matrix is decoupled from the equations in the fractures. Thus, the dimensionless advective velocity in the fracture is~$\mathcal K$. In Figure~\ref{fig:hfrac_F}, cases with equal permeability ratio have the same line type, and in the figure it can be seen that the mixing length for small $\hat t$ is independent of $\mathcal A$ and only vary with the permeability ratio~$\mathcal K$. %
When the concentration front in the fracture outgrows that in the rock matrix, two fluxes occur. First, there's a diffusive flux from the forward finger in the fracture to the rock matrix due to the concentration gradient. Second, an advective flux arises from pressure differences—higher pressure in the fracture ahead of the concentration front and lower pressure just behind it, driven by the viscosity ratio. This pressure difference increases as the finger lengthens, reducing the growth rate of the mixing length, as depicted in Figure~\ref{fig:hfrac_F}. The advective flux from a leading finger is also observed in studies by~\citet{budeck2015}, which investigated viscous fingering in connected microchannels. The mixing length's growth rate slows until it reaches equilibrium between the fracture's volumetric flux and exchange with the rock matrix. Figure~\ref{fig:hfrac_F} illustrates this as the convergence of mixing length in simulations with the same volumetric flux ratio, $\mathcal{KA}$ (lines of the same color).

From the mixing length and the number of fingers plotted in Figures~\ref{fig:hfrac_KvsA} and~\ref{fig:hfrac_F} we conclude
that increasing the permeability ratio, $\mathcal K$, or the dimensionless aperture, $\mathcal A$, increases the mixing length and decreases the number of fingers by up to several orders of magnitude. The effect of the fractures on the quantitative measures is largest before the onset of the viscous fingering in the homogeneous case, $\hat t \sim 1$, for the current parameters. At early times, the mixing length is only dependent on the permeability ratio, but when the fracture interacts with the rock matrix the behavior of the flow changes and the governing parameter of the fracture is the volumetric flux ratio $\mathcal{KA}$. Thus, at intermediate to late stages the permeability contrast can vary three orders of magnitude but will not significantly affect the qualitative or quantitative measures if the volumetric flux ratio is fixed.

\subsubsection{Influence of the fracture density}\label{sec:parallel_N}
In the previous subsection, the width between the fractures was equal to the
characteristic width $H$. In this section we increase the fracture density in
the domain, which leaves less space for viscous fingers within the rock
matrix. In all simulations in this subsection, the parameters
$\{\Pen, R, \mathcal K, \mathcal A\} = \{1000, 2, 10, 10^{-3}\}$ are fixed, and
the fracture density is varied from $1$ to $5$. The fractures are evenly spaced
in the domain.
\begin{figure}[h]
  \centering \def\figWidth{0.49\textwidth}
 \begin{minipage}[c]{.97\textwidth}
  \begin{subfigure}[b]{\figWidth}
    \centering%
    \def\tikzWidth{0.9\textwidth}
    \begin{tikzpicture}
      \node[inner sep=2ex, anchor=south east] (00) at (0,0)
      {\includegraphics[width=\tikzWidth]{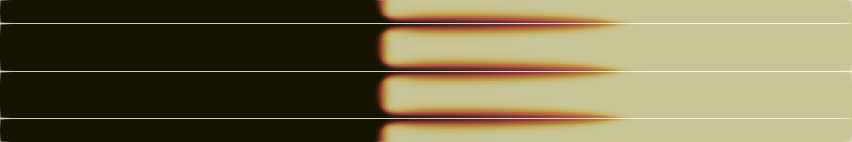}};
      \node[anchor=north, yshift=2.5ex, xshift=2.5ex] at (00.south west) {-2};
      \node[anchor=north, yshift=2.5ex] (mid) at (00.south) {1};%
      \node[anchor=north, yshift=2.5ex, xshift=-3ex] at (00.south east) {4};
      \node[anchor=east, yshift=2.5ex, xshift=2.5ex] at (00.south west) {0};
      \node[anchor=east, yshift=2.5ex, xshift=2.5ex] at (00.west) {1};
      \node[anchor=east] at (00.west) {$\hat y$}; \node[anchor=north, yshift=1ex]
      at (mid.south) {$\hat x$};
    \end{tikzpicture}
    \caption{ $\hat t = 1.0$}
  \end{subfigure}%
  \begin{subfigure}[b]{\figWidth}
    \centering%
    \def\tikzWidth{0.9\textwidth}
    \begin{tikzpicture}
      \node[inner sep=2ex, anchor=south east] (00) at (0,0)
      {\includegraphics[width=\tikzWidth]{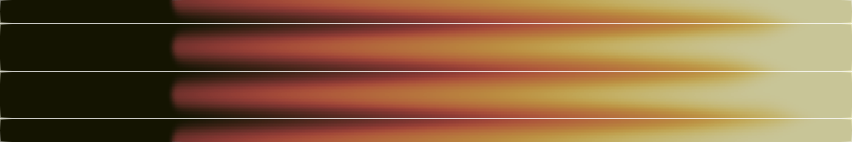}};
      \node[anchor=north, yshift=2.5ex, xshift=2.5ex] at (00.south west) {4};
      \node[anchor=north, yshift=2.5ex] (mid) at (00.south) {10};%
      \node[anchor=north, yshift=2.5ex, xshift=-3ex] at (00.south east) {16};
      \node[anchor=east, yshift=2.5ex, xshift=2.5ex] at (00.south west) {0};
      \node[anchor=east, yshift=2.5ex, xshift=2.5ex] at (00.west) {1};
      \node[anchor=north, yshift=1ex] at (mid.south) {$\hat x$};
    \end{tikzpicture}
    \caption{ $\hat t = 10$}
  \end{subfigure}
  \begin{subfigure}[b]{\figWidth}
    \centering%
    \def\tikzWidth{0.9\textwidth}
    \begin{tikzpicture}
      \node[inner sep=2ex, anchor=south east] (00) at (0,0)
      {\includegraphics[width=\tikzWidth]{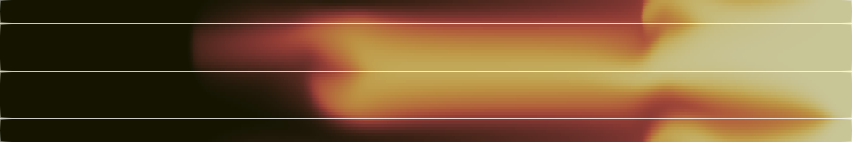}};
      \node[anchor=north, yshift=2.5ex, xshift=2.5ex] at (00.south west) {40};
      \node[anchor=north, yshift=2.5ex] (mid) at (00.south) {50};%
      \node[anchor=north, yshift=2.5ex, xshift=-3ex] at (00.south east) {60};
      \node[anchor=east, yshift=2.5ex, xshift=2.5ex] at (00.south west) {0};
      \node[anchor=east, yshift=2.5ex, xshift=2.5ex] at (00.west) {1};
      \node[anchor=east] at (00.west) {$\hat y$}; \node[anchor=north, yshift=1ex]
      at (mid.south) {$\hat x$};
    \end{tikzpicture}
    \caption{$\hat t = 50$}
  \end{subfigure}%
  \begin{subfigure}[b]{\figWidth}
    \centering%
    \def\tikzWidth{0.9\textwidth}
    \begin{tikzpicture}
      \node[inner sep=2ex, anchor=south east] (00) at (0,0)
      {\includegraphics[width=\tikzWidth]{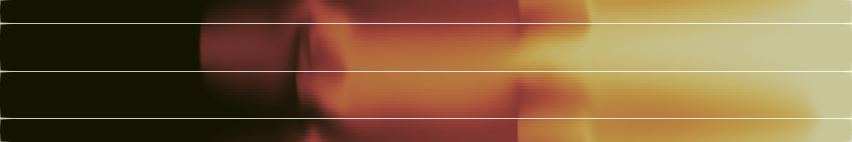}};
      \node[anchor=north, yshift=2.5ex, xshift=2.5ex] at (00.south west) {75};
      \node[anchor=north, yshift=2.5ex] (mid) at (00.south) {100};%
      \node[anchor=north, yshift=2.5ex, xshift=-3ex] at (00.south east) {125};
      \node[anchor=east, yshift=2.5ex, xshift=2.5ex] at (00.south west) {0};
      \node[anchor=east, yshift=2.5ex, xshift=2.5ex] at (00.west) {1};
      \node[anchor=north, yshift=1ex] at (mid.south) {$\hat x$};
    \end{tikzpicture}
    \caption{$\hat t = 100$}
  \end{subfigure}%
  \end{minipage}
\begin{minipage}[c]{.02\textwidth}
  \begin{subfigure}[b]{\textwidth}
  \includegraphics[width=\textwidth]{fig/colorbar.png}
  \end{subfigure}%
  \end{minipage}%
  \\%
  \hspace{\legendWidth}%
  \begin{subfigure}[T]{0.7\textwidth}
    \centering \includegraphics[width=1\textwidth]{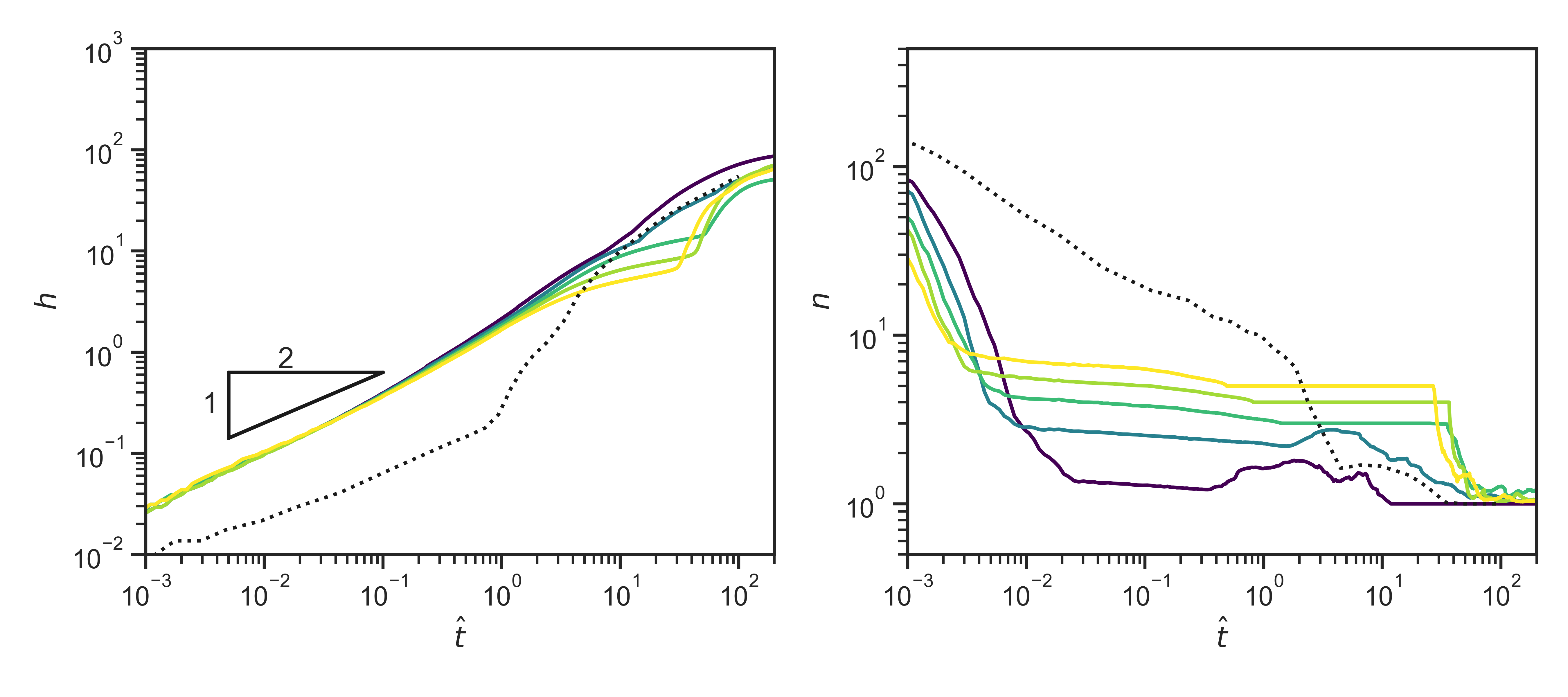}
    \caption{}
  \end{subfigure}  
  \hspace{-1em}
  \begin{subfigure}[T]{\legendWidth}
    \vspace{1em}
    \includegraphics[width=1\textwidth]{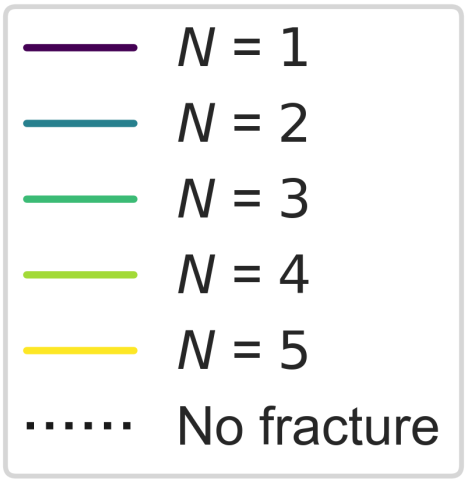}
    \end{subfigure}
  \caption{Evolution of the concentration field for the case with fractures parallel to the principle flow directions. The parameters used in the simulations are
    $\{\Pen, R, \mathcal K, \mathcal A\} = \{1000, 2, 10, 10^{-3}\}$. (a) - (d)
    Color maps of the concentration field for $N = 3$ at different times. Note the
    increasing aspect ratio in the figures.  (e) The mixing length and number of
    fingers for different fracture density. The dotted lines show the
    evolution for the homogeneous case without fractures.}
  \label{fig:hfrac_N}
\end{figure}

Figure~\ref{fig:hfrac_N} shows the time evolution of the concentration field for
the case $N=3$. All other values of $N$ show the same qualitative behaviour where the flow goes through three distinct regimes. In the first regime there are stable propagating fingers aligned with the fractures. In the second regime there is viscous fingering across the whole domain, and in the final regime there is a single propagating finger. In the following we discuss each regime in detail.

Initially, a finger forms around each fracture, which quickly outgrows the diffusive front.
Mass exchange from fracture to matrix then slows finger growth, leading to close-to-square-root mixing length growth, as seen in Figure~\ref{fig:hfrac_N}.
For the cases with $N=1$ and $N=2$, fractures are spaced sufficiently apart for viscous fingers to develop in the rock matrix between fractures, slightly increasing the number of fingers around $\hat t\sim 1$. When the viscous fingering starts in the rock matrix the fingers aligned with the fractures are disturbed and the viscous fingering behaves as for an homogeneous medium. For the cases with $N \ge 3$ no viscous fingers
are observed in the rock matrix before the second regime characterized by the viscous fingering across the fracture geometry. This transition occurs when the previous stable displacement with one finger coinciding with each fracture becomes unstable, and the viscous fingers interact across the fractures. This is evident in Figure~\ref{fig:hfrac_N} as a sudden drop in the number of fingers and an increase in mixing length growth rate. The continued evolution of the new viscous fingering behaves similar to the viscous fingering in the homogeneous case, with tip splitting, shielding, and finger coarsening until only one finger is left in the domain. This behaviour resembles observations in layered porous media~\citep{nijjer2019stable}, characterized by viscous fingering across highly permeable layers.

\subsubsection{Influence of the P{\' e}clet number and the log viscosity ratio}
\label{sec:hfrac_PeR}
To study how the fractured domain is affected by the P{\' e}clet number and log
viscosity ratio, we consider a domain with a single horizontal fracture that is
centered in the domain. The permeability ratio and dimensionless aperture are
fixed to $\mathcal K = 10$ and $\mathcal A = 10^{-3}$. Two sets of simulations are
run. In the first set, the P{\' e}clet number is varied,
$\Pen = 100, 500, 1000, 2000, 4000$, and the log viscosity ratio is fixed to $R = 2$.
In the second set, the log viscosity ratio is varied, $R = 1, 2, 3, 4$, and the
P{\' e}clet number is fixed to 1000.
\begin{figure}[h]
  \centering%
  \includegraphics[width=1\textwidth]{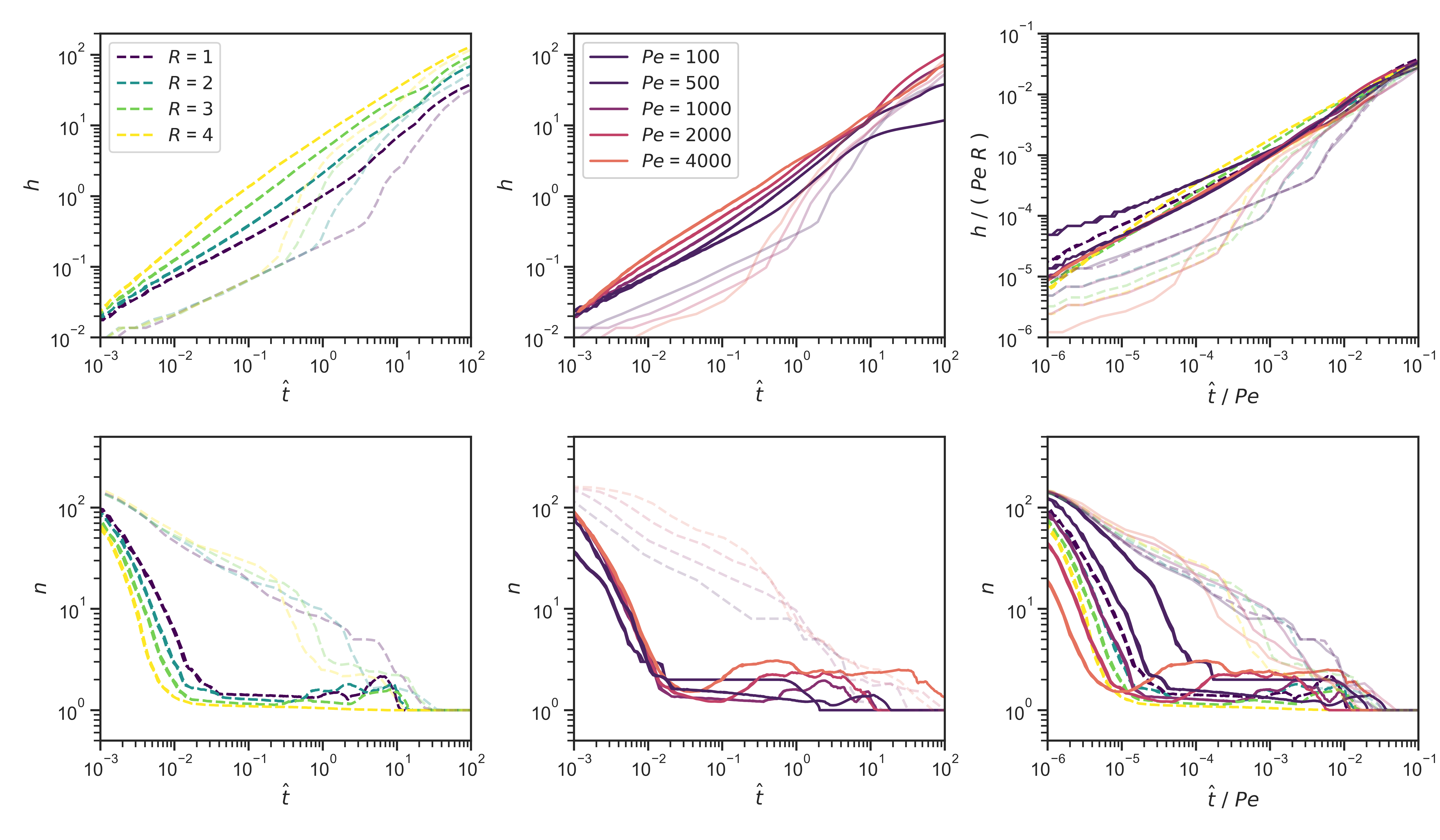}\vspace{-1em}
  \caption{The mixing length (top row) and the number of fingers (bottom row) for
    the test case with fractures parallel to the principal flow direction. The parameters
    $\{\mathcal K, \mathcal A, N\} = \{10, 10^{-3}, 1\}$ are fixed. The transparent
    lines show the values from simulations without fractures. Two sets of
    simulations are run. In the first set (left column) the log viscosity ratio is
    varied while $\Pen = 1000$ is fixed. In the second set (middle column), the P{\' e}clet number is
    varied, while $R=2$ is fixed. The right column shows all simulations with rescaled length and time.}
  \label{fig:hfrac_PeR}
\end{figure}

Figure~\ref{fig:hfrac_PeR} shows the mixing length and number of fingers for
all the cases. As we have seen in Section~\ref{sec:paralell_KA}, a single finger
forms along the fracture and this finger causes the early time ($t\sim10^{-3}$) difference in mixing length. At this time both the mixing length and number of fingers are independent of the P{\' e}clet number and log viscosity ratio. For times $10^{-2}\lesssim \hat t\lesssim10^1$ increasing the log viscosity ratio or the P{\' e}clet number increases the mixing length. The onset of viscous fingers in the rock matrix between the fractures can be
seen as the increase of the number of fingers. We observe more viscous fingers and an earlier onset time in the matrix for higher P{\' e}clet numbers.
\enlargethispage{1\baselineskip}

After the onset of the viscous fingering in the homogeneous case at $\hat t\sim 1$, the mixing length in the homogeneous domain grows linearly and faster than the fractured case until they coincide. Eventually, in the final stages of all simulations, both fractured and homogeneous cases exhibit a single propagating finger, behaving qualitatively and quantitatively as in a homogeneous domain. \citet{nijjer_2018} shows that for late times, $\hat t\sim \Pen$, the mixing length for a homogeneous domain scales as $h\sim R\Pen$.
In the right column of Figure~\ref{fig:hfrac_PeR} we apply this scaling to the fracture case and show that after the shutdown of the viscous fingering, the mixing length and the number of fingers collapses onto this late time scaling. This is equivalent to the observation in Sections~\ref{sec:paralell_KA} and~\ref{sec:parallel_N} where simulations with dimensionless volumetric flux ratios $\mathcal K\mathcal A\le 10^{-2}$ approached the behavior of the homogeneous case for times $\hat t\sim \Pen$. 

From the simulations in this subsection with a permeability ratio, $\mathcal K = 10^1$, and dimensionless aperture, $\mathcal A = 10^{-3}$, we can conclude that the influence of the fractures is most important up to the onset of the viscous fingers, and the effect of the fracture is negligible at late times,
$\hat t\sim \Pen$.

\subsection{Brick shaped fracture networks}\label{sec:brick}
In the previous subsection we learned important lessons about the importance of the volumetric flux ratio when describing the effects fractures have on viscous fingering. In this section we consider variants of the brick shaped fracture network depicted in Figure~\ref{fig:fracture_networks_brick} to investigate further how the permeability ratio and dimensionless aperture affect the flow paths for a more complex geometry. The length of the bricks in $\hat x$-direction is twice the width of the bricks in $\hat y$-direction. 

\subsubsection{Influence of the permeability ratio and dimensionless aperture}
\label{sec:brick_KA}
\begin{figure}
  \centering \def\figWidth{0.3\textwidth}
  \includegraphics[width=1\textwidth]{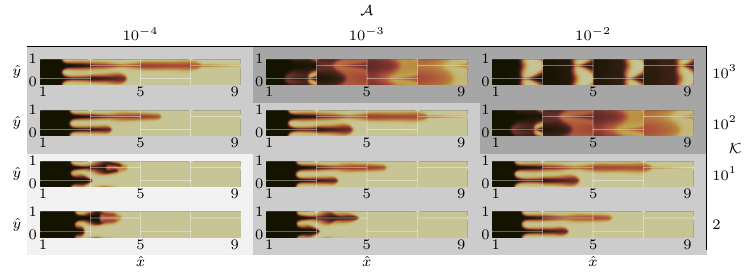}
  \includegraphics[width=0.2\textwidth]{fig/colorbar_h.png}
  \caption{Domain with a brick pattern, the fractures are shown as white lines. The
    figure shows the concentration for $\hat t=3$, for different permeability
    ratios and dimensionless apertures. The other parameters are fixed to
    $\{R, \Pen , N\} = \{2, 500, 1\}$. The different background shades of gray indicate different qualitative regimes. Note that the simulation domains are much
    larger than shown in the figures. \label{fig:brick_KvsA_snap}}
\end{figure}%

\begin{figure}
  \centering \def\figWidth{1\textwidth}
  \begin{subfigure}[T]{0.8\figWidth}
    \centering \includegraphics[width=1\textwidth]{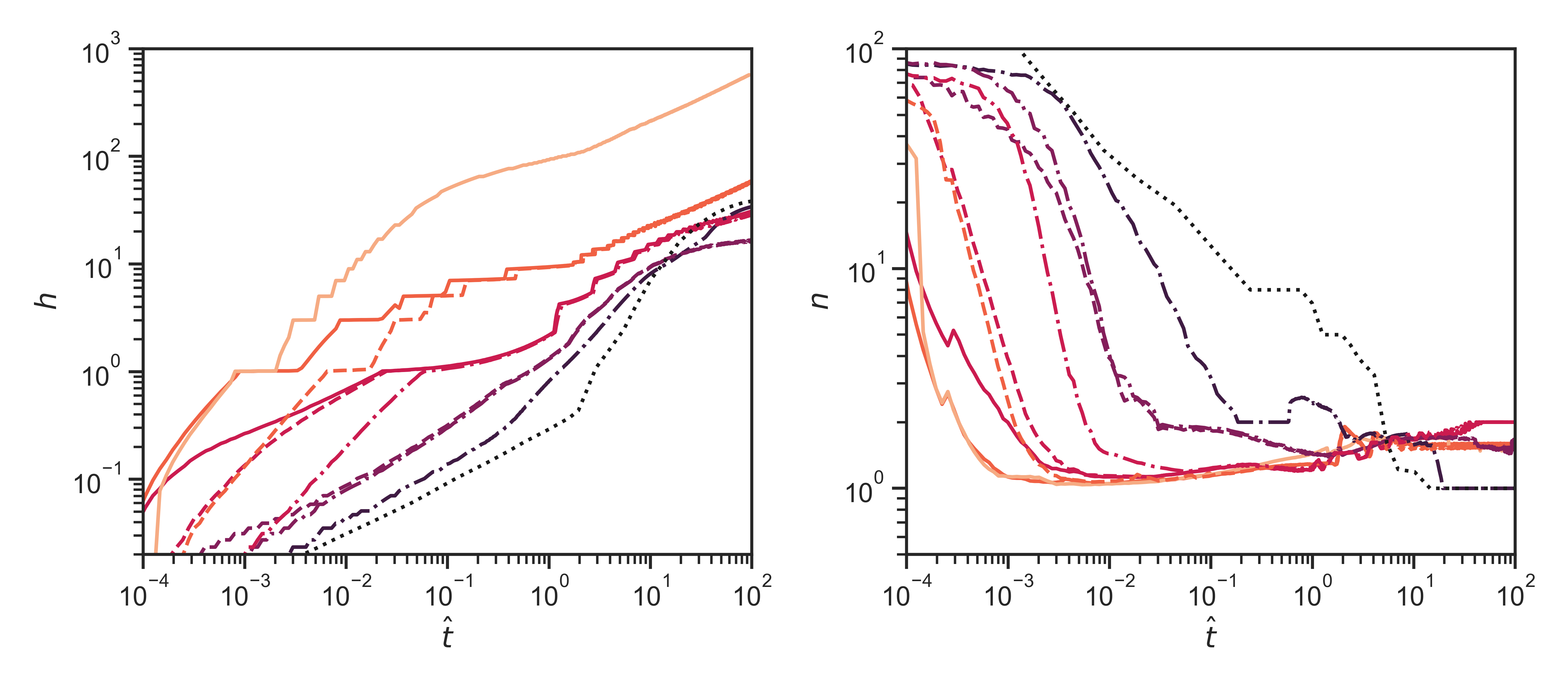}
  \end{subfigure}
  \hspace{-1em}
  \begin{subfigure}[T]{\legendWidth}
    \vspace{1em}
    \includegraphics[width=1\textwidth]{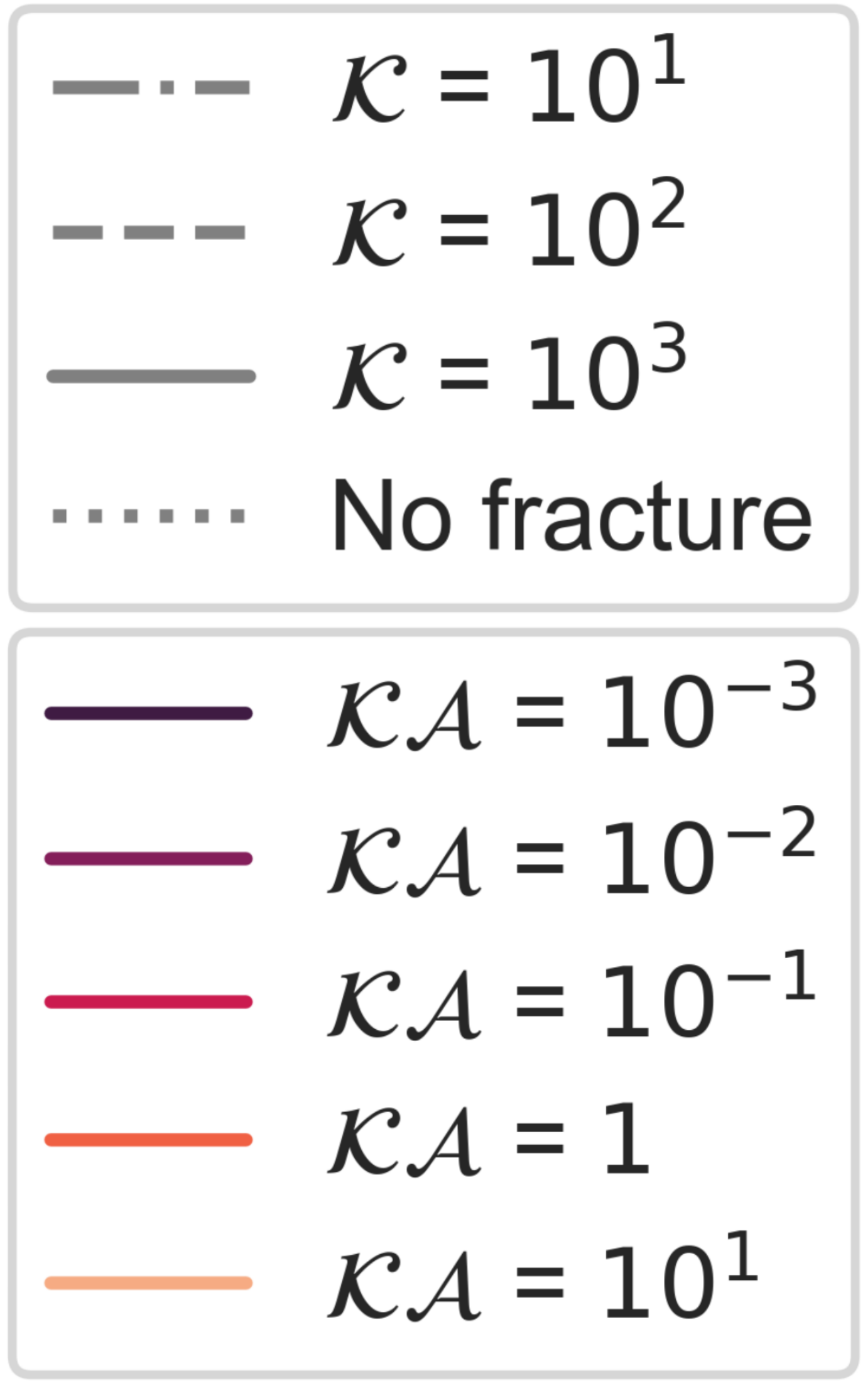}
    \end{subfigure}
  \caption{Evolution of the concentration field for the brick shaped fracture network.
    The other parameters are fixed to $\{Pe, R, N\} = \{500, 2, 1\}$ for varying $\mathcal K$ and $\mathcal{KA}$. The dotted black
    line is a simulation without any fractures. The colors show the volumetric
    flux ratio $\mathcal K\mathcal A$, and the line style the value of $\mathcal
    K$. The left figure shows the mixing length and the right figure shows the number of fingers.
  }
  \label{fig:brick_F}
\end{figure}
In the simulations in this section, the P{\' e}clet number is fixed to $\Pen = 500$, the log viscosity ratio to $R = 2$, and $N=1$. Note that $N$ is the fracture density in a vertical slice of the domain, thus the brick network has two sets
of discontinuous horizontal lines as seen in Figure~\ref{fig:brick_KvsA_snap}.

The general trend in the simulations is equivalent for all simulations. A finger
develops around the horizontal fracture, and when the horizontal fracture ends,
the finger continues through the rock matrix until it reaches the next
horizontal fracture. The exception is for the highest volumetric flux ratios
$\mathcal K\mathcal A \ge 10^0$ where the fluid continues
through the vertical fractures when the horizontal fractures end. This can be observed in Figure~\ref{fig:brick_KvsA_snap}, displaying snapshots of the concentration profile at
$\hat t = 3$ for all simulations. The first observation we make is that there
are almost no viscous fingers in any of the parameter combinations. Only for the smallest volumetric flux ratios do a few fingers emerge in the rock matrix between the fractures (lightest shade in the lower-left corner of the figure). The vertical fractures are perpendicular to the principle flow direction which causes the intricate zig-zag patterns in the rock matrix domain for the volumetric flux ratios $\mathcal{KA}\ge 1$.

It is striking how simulations with fixed volumetric flux ratio
$\mathcal K \mathcal A$ (diagonals in Figure~\ref{fig:brick_KvsA_snap}) are
remarkable similar, even for cases with three orders of magnitude difference in
permeability ratio. The quantitative measures for the same volumetric flux ratio also collapse at intermediate to late times, as can be seen in Figure~\ref{fig:brick_F}. The dependence on the volumetric flux ratio is equivalent to the
behaviour observed for the parallel fractures in Section~\ref{sec:parallel}. At early times, the flow is determined by the permeability ratio (lines with the same line style in Figure~\ref{fig:brick_F}), however, as the viscous fingers grow, the
dynamics of the matrix-fracture interaction changes and becomes independent of the permeability contrast.  At times $\hat t >0.1$ it is the volumetric flux ratio $\mathcal K \mathcal A$ that governs the fluid flow for the parameter range in this study, but the exact transition time vary between simulations. This is shown in Figure~\ref{fig:brick_F} where it can be seen that simulations with equal volumetric flux ratio $\mathcal K \mathcal A$ (lines with same color) have an almost equal mixing length and number of fingers.

Finally, we observe that due to the zig-zag pattern of the fracture network, the mixing length for the highest permeability ratios (seen in Figure~\ref{fig:brick_F}) is reduced by up to an order of magnitude compared to the parallel fractures (seen in Figure~\ref{fig:hfrac_F}). The fluid flow also has a different qualitative behaviour as seen by comparing the darkest regions (upper right corner) of Figures~\ref{fig:parallel_KvsA_snap} and~\ref{fig:brick_KvsA_snap}.

\subsubsection{Influence of the fracture density}
In the previous subsection, the transversal distance between the fractures was equal the
characteristic width $H$. In this section we increase the fracture density in
the domain, which leaves less space for viscous fingers within the rock
matrix. In all simulations in this section the parameters
$\{\Pen, R, \mathcal K, \mathcal A\} = \{500, 2, 10, 10^{-3}\}$ are fixed, and the
fracture density is varied from $N = 1$ to $N = 5$.

\begin{figure}[h]
  \centering%
  \def\figWidth{0.49\textwidth}%
  \def\figHeight{0.16667\textwidth}%
  \def\tikzWidth{0.9\textwidth}
  \begin{minipage}{0.97\textwidth}
  \begin{subfigure}[b]{\figWidth}
    \centering
    \begin{tikzpicture}
      \node[inner sep=2ex, anchor=south east] (00) at (0,0) {
        \includegraphics[height=\figHeight,width=\tikzWidth]{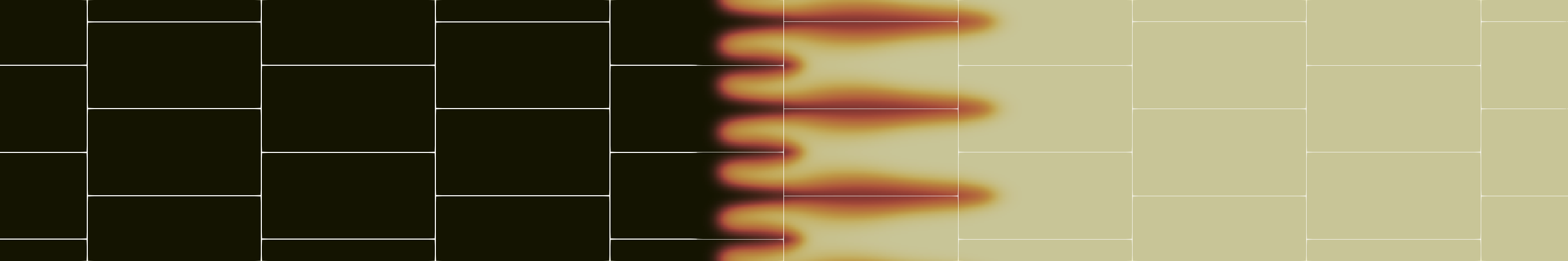}};
      \node[anchor=north, yshift=2.5ex, xshift=2.5ex] at (00.south west) {-2};
      \node[anchor=north, yshift=2.5ex] at (00.south) {1}; \node[anchor=north,
      yshift=2.5ex, xshift=-3ex] at (00.south east) {4}; \node[anchor=south,
      yshift=-1.2em] at (00.south) {$\hat x$};
      \node[anchor=east, yshift=2.5ex, xshift=2.5ex] at (00.south west) {0};
      \node[anchor=east, yshift=-2.5ex, xshift=2.5ex] at (00.north west) {1};
      \node[anchor=east, xshift=1ex] at (00.west) {$\hat y$};
    \end{tikzpicture}
    \caption{ $\hat t = 1$}
  \end{subfigure}%
  \begin{subfigure}[b]{\figWidth}
    \centering
    \begin{tikzpicture}
      \node[inner sep=2ex, anchor=south east] (00) at (0,0) {
        \includegraphics[height=\figHeight,width=\tikzWidth]{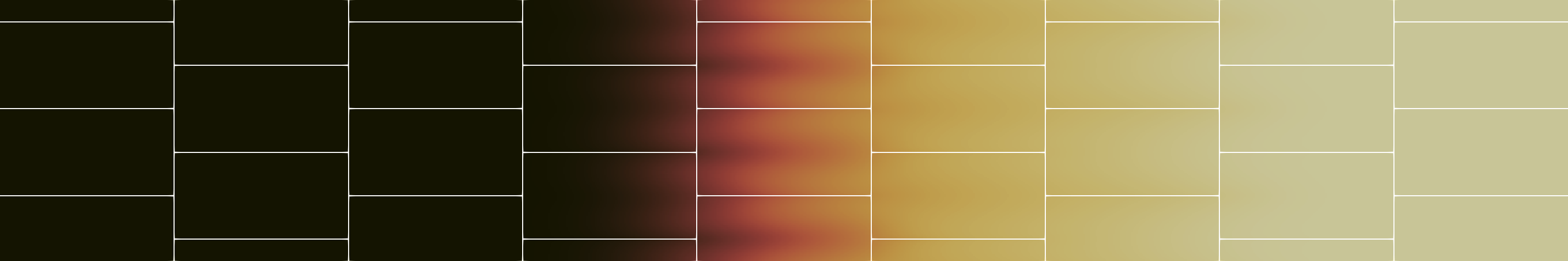}};
      \node[anchor=north, yshift=2.5ex, xshift=2.5ex] at (00.south west) {5};
      \node[anchor=north, yshift=2.5ex] at (00.south) {8}; \node[anchor=north,
      yshift=2.5ex, xshift=-3ex] at (00.south east) {11}; \node[anchor=south,
      yshift=-1.2em] at (00.south) {$\hat x$};
      \node[anchor=east, yshift=2.5ex, xshift=2.5ex] at (00.south west) {0};
      \node[anchor=east, yshift=-2.5ex, xshift=2.5ex] at (00.north west) {1};
      \node[anchor=east, xshift=1ex] at (00.west) {$\hat y$};
    \end{tikzpicture}
    \caption{ $\hat t = 8$}
  \end{subfigure}
  \begin{subfigure}[b]{\figWidth}
    \centering
    \begin{tikzpicture}
      \node[inner sep=2ex, anchor=south east] (00) at (0,0) {
        \includegraphics[height=\figHeight,width=\tikzWidth]{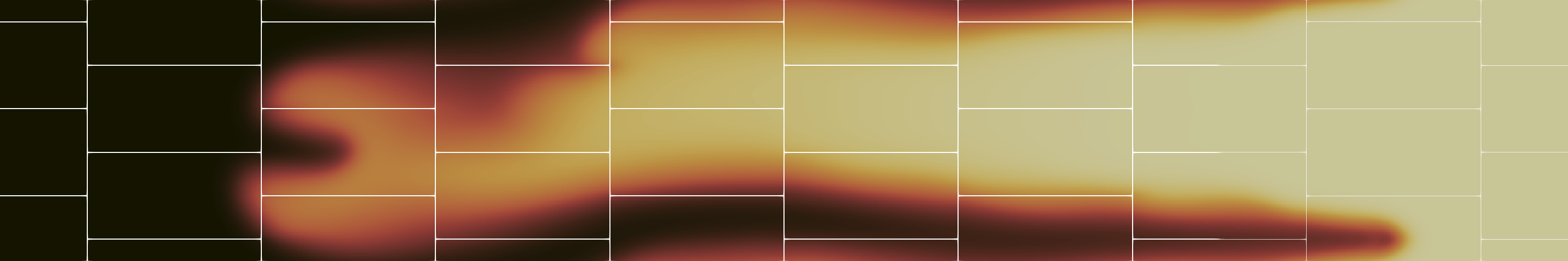}};
      \node[anchor=north, yshift=2.5ex, xshift=2.5ex] at (00.south west) {12};
      \node[anchor=north, yshift=2.5ex] at (00.south) {15}; \node[anchor=north,
      yshift=2.5ex, xshift=-3ex] at (00.south east) {18}; \node[anchor=south,
      yshift=-1.2em] at (00.south) {$\hat x$};
      \node[anchor=east, yshift=2.5ex, xshift=2.5ex] at (00.south west) {0};
      \node[anchor=east, yshift=-2.5ex, xshift=2.5ex] at (00.north west) {1};
      \node[anchor=east, xshift=1ex] at (00.west) {$\hat y$};
    \end{tikzpicture}
    \caption{$\hat t = 15$}
  \end{subfigure}%
  \begin{subfigure}[b]{\figWidth}
    \centering
    \begin{tikzpicture}
      \node[inner sep=2ex, anchor=south east] (00) at (0,0) {
        \includegraphics[height=\figHeight,width=\tikzWidth]{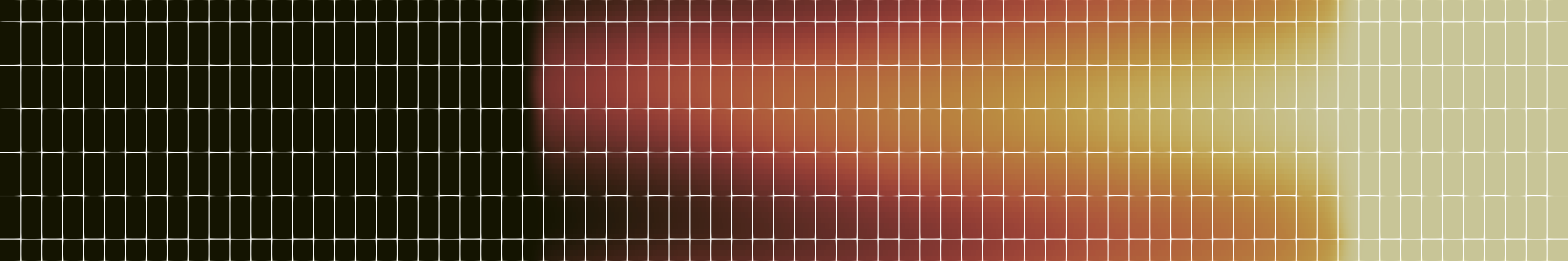}};
      \node[anchor=north, yshift=2.5ex, xshift=2.5ex] at (00.south west) {25};
      \node[anchor=north, yshift=2.5ex] at (00.south) {50}; \node[anchor=north,
      yshift=2.5ex, xshift=-3ex] at (00.south east) {75}; \node[anchor=south,
      yshift=-1.2em] at (00.south) {$\hat x$};
      \node[anchor=east, yshift=2.5ex, xshift=2.5ex] at (00.south west) {0};
      \node[anchor=east, yshift=-2.5ex, xshift=2.5ex] at (00.north west) {1};
      \node[anchor=east, xshift=1ex] at (00.west) {$\hat y$};
    \end{tikzpicture}
    \caption{$\hat t = 50$}
  \end{subfigure}%
  \end{minipage}%
  \hfill%
  \begin{minipage}[c]{.02\textwidth}
  \begin{subfigure}[b]{\textwidth}
  \includegraphics[width=\textwidth]{fig/colorbar.png}
  \end{subfigure}%
  \end{minipage}%
  \\%
  \noindent
  \hspace{\legendWidth}
  \begin{subfigure}[T]{0.7\textwidth}
    \centering
    \includegraphics[width=1\textwidth]{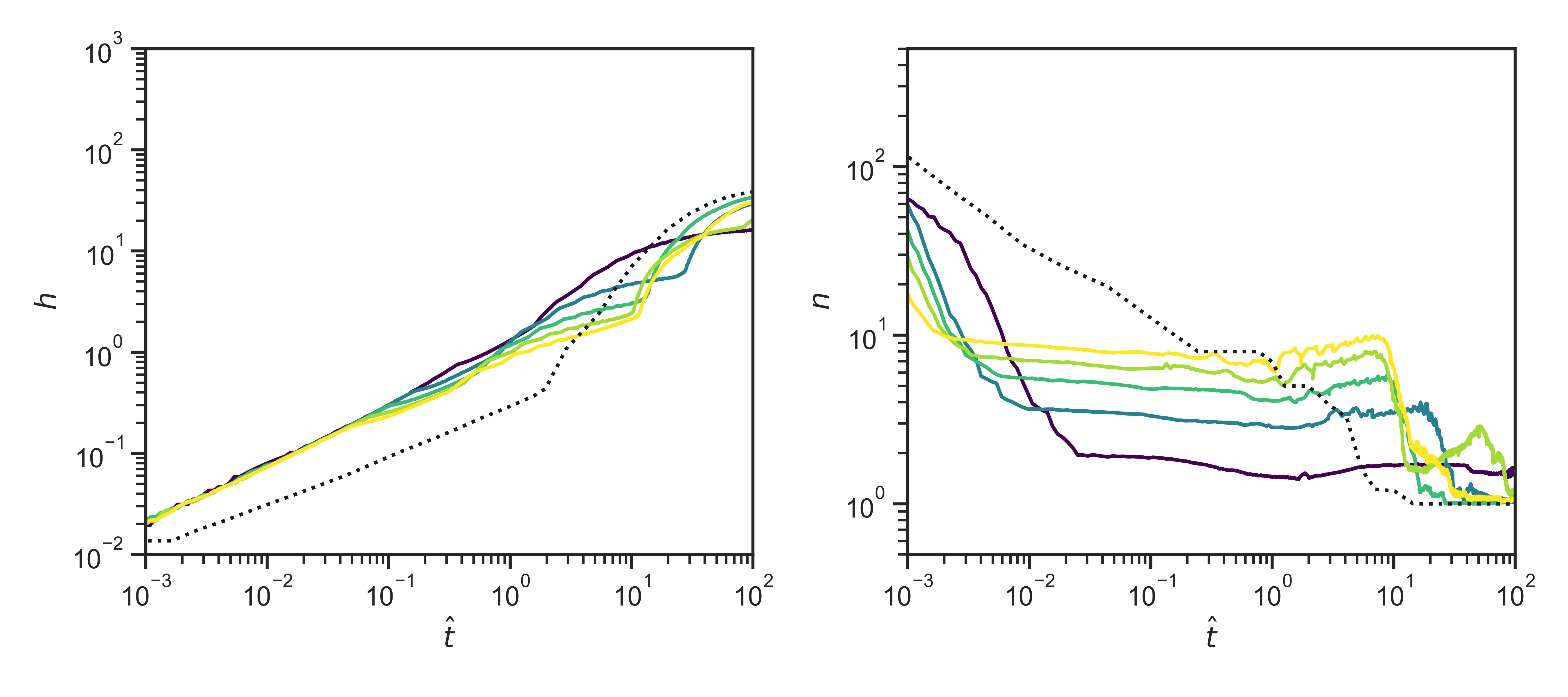}
    \caption{}
  \end{subfigure}
  \hspace{-1em}
  \begin{subfigure}[T]{\legendWidth}
    \vspace{1em}
    \includegraphics[width=1\textwidth]{fig/parallel/h_frac_vs_N_legend.png}
    \end{subfigure}
  \caption{Evolution of the concentration field for the brick shaped fractures. The parameters
    $\{\Pen, R, \mathcal K, \mathcal A\} = \{500, 2, 10, 10^{-3}\}$ are fixed, while the fracture density is varied. (a) - (d)
    Color maps of the concentration field for $N=3$ at different times. Note the increasing aspect ratio in the figures. (e) The
    mixing length and number of fingers for different fracture densities.}
  \label{fig:brick_N}
\end{figure}

Figure~\ref{fig:brick_N} shows the time evolution of the case $N = 3$. The other
simulations behave in a qualitative similar manner. In the simulation, a finger is initiated
along each fracture as can be seen in subfigure (a). When the horizontal fractures
end, the fingers continue through the rock matrix instead of following the
fracture network around the block. When the trailing displacement front in the
rock matrix reaches the next set of bricks located at $\hat x = 0.5$, a new set of
fingers is initiated along the second set of horizontal fractures. The new set of
fingers and old set of fingers (six in total) continue to grow and we do not observe any other fingers in the domain and the displacement is stable with the six propagating fingers growing longer (as seen in subfigure (b)). At times $\hat t \approx 10$ the fingers that are aligned with the fractures become unstable and start to compete. The viscous fingers then behaves similar to the viscous instabilities in the homogeneous case with merging, tip splitting and shielding until all fingers have merge to a single finger (c).
The merging of the fingers is clearly seen in subfigure (e) as the sudden drop in the number of fingers at time $\hat t\approx 20$ and as an increase in the growth rate of the mixing length $h$.

\subsection{Random fracture networks}
The random fracture networks are generated by creating two sets of fractures, one horizontal set and one vertical set. The fractures in both sets are generated from two random variables, the coordinate of the starting point of the fractures and the fracture length. The starting point follows a uniform distribution in the domain. The fracture lengths are uniformly distributed in the intervals $[0, 5]$ and $[0, 1]$ in $\hat x$- and $\hat y$-direction, respectively. Due to the periodicity of the boundary, the vertical fractures with a starting point $\hat y_s$ and length $\Delta \hat y > 1-\hat y_s$ are wrapped around the periodic boundary and extend a length $\Delta \hat y - (1-\hat y_s)$ from the bottom boundary. To avoid arbitrary close fractures, all fractures are snapped to a coarse Cartesian grid with dimensionless resolution 0.1. 

\subsubsection{Influence of the volumetric flux ratio}\label{sec:random_F}
To study the influence of the volumetric flux ratio, a suite of simulations is set up with the parameters $\{\Pen, R, \mathcal A\} = \{500, 2, 10^{-3}\}$ fixed while the permeability ratio and fracture density are varied. For the random network, significant variations in the quantitative measures are observed among each realization of the fracture network. To account for this variability, five simulations were conducted for each parameter set. 

\begin{figure}
  \centering \def\figWidth{0.3\textwidth}
  \includegraphics[width=1\textwidth]{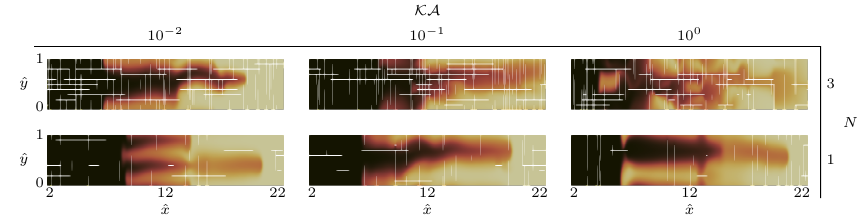}
  \includegraphics[width=0.2\textwidth]{fig/colorbar_h.png}
  \caption{Domain with a random fracture network, the fractures are shown as white lines. The figure shows the concentration for $\hat t=12$, for different volumetric flux ratios and fracture density. The other parameters are fixed to $\{R, \Pen , \mathcal A\} = \{2, 500, 10^{-3}\}$. Note that the simulation domains are much
    larger than shown in the figures. \label{fig:random_KA_evo}}
\end{figure}

\begin{figure}[h]
  \centering%
  \begin{minipage}[t]{0.7\textwidth}
      \vspace{0pt}
      \begin{subfigure}[t]{\textwidth}
        \centering%
        \includegraphics[width=\textwidth]{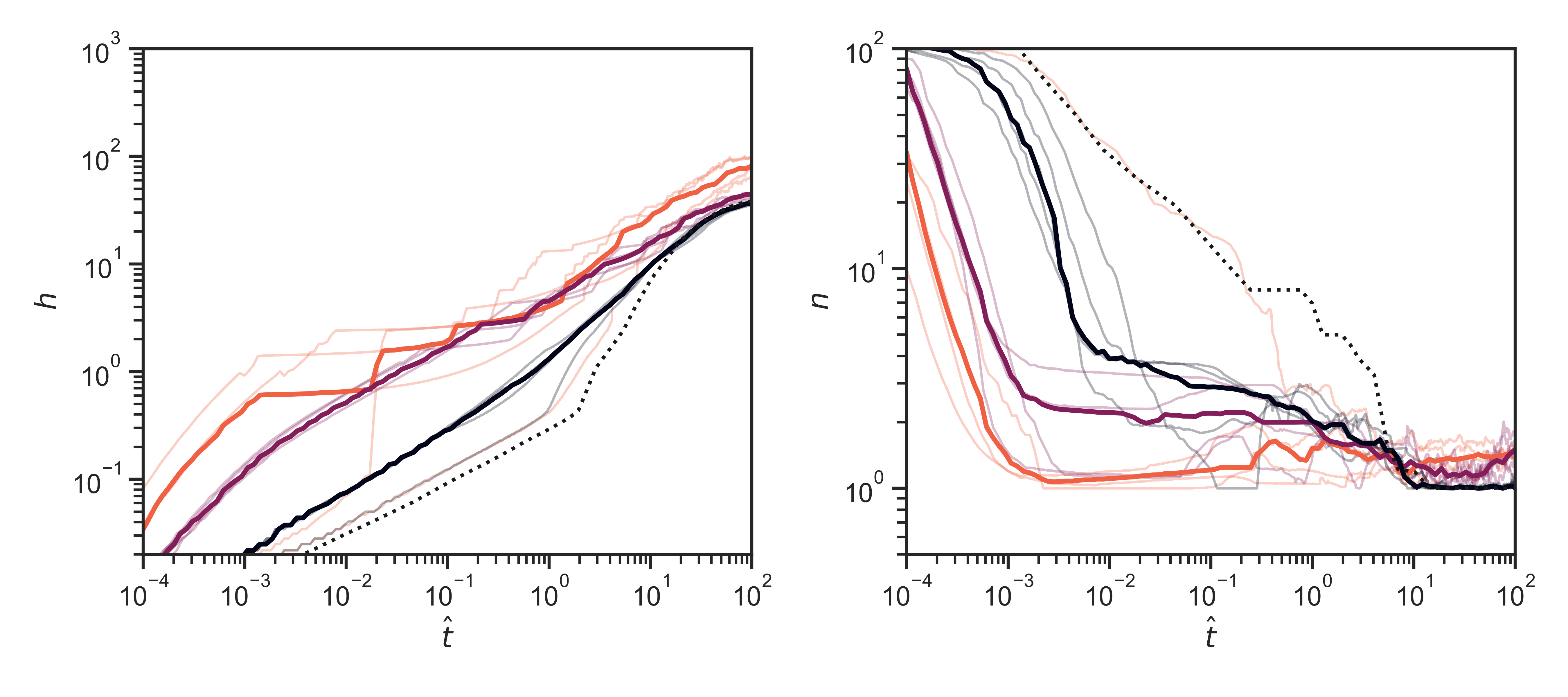}
        \caption{$N = 1$}
      \end{subfigure}
      \begin{subfigure}[t]{\textwidth}
        \centering
        \includegraphics[width=1\textwidth]{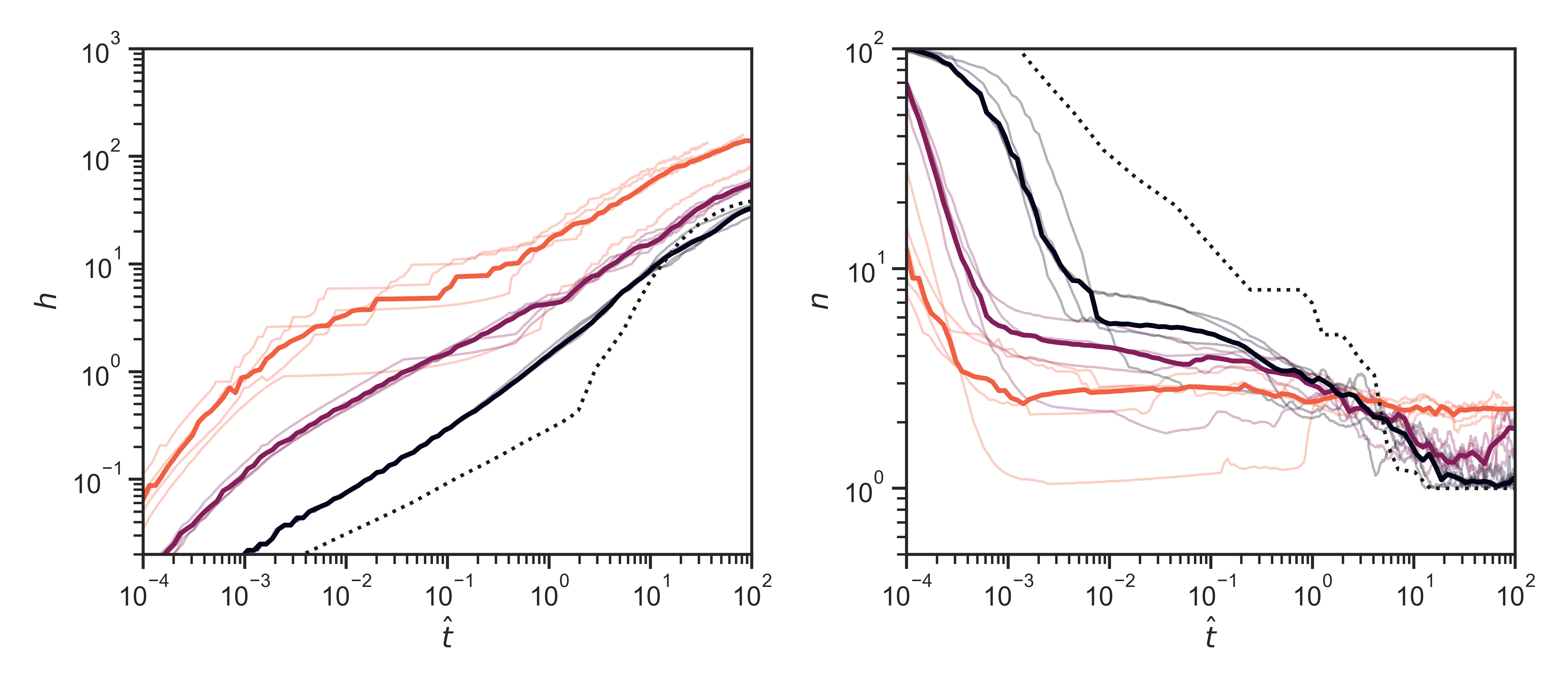}
        \caption{$N=3$}
      \end{subfigure}
  \end{minipage}
  \hspace{-1em}
  \begin{minipage}[t]{\legendWidth}
    \vspace{1em}
    \includegraphics[width=1\textwidth]{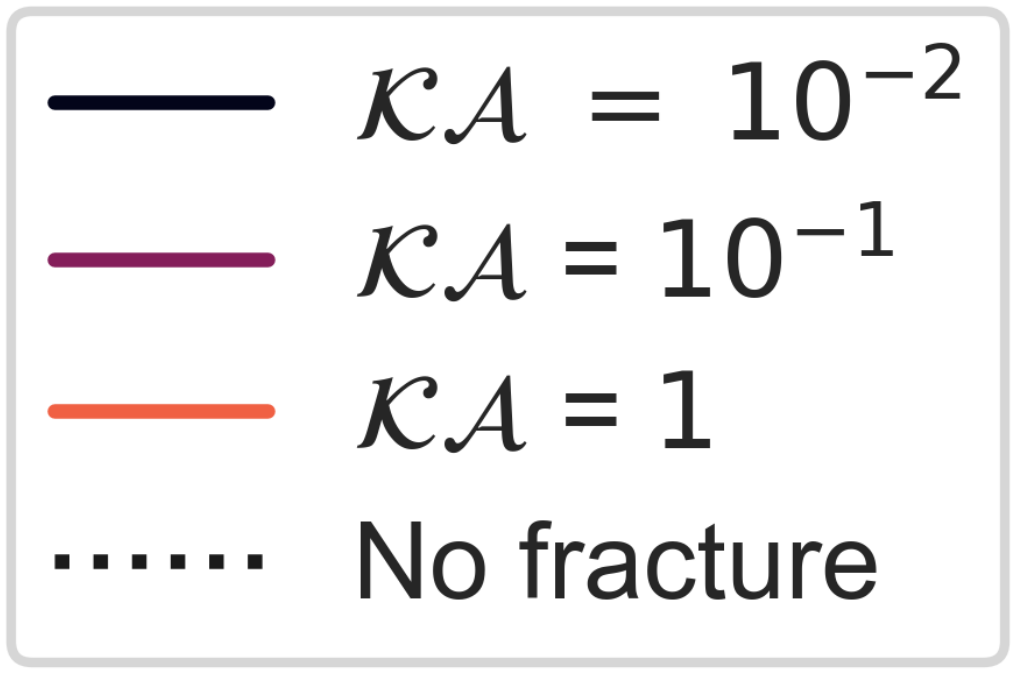}
    \end{minipage}
  \caption{Finger length $h$ and number of fingers $n$ for varying $\mathcal{KA}$ and $N$ for the random fracture network. 
  The solid lines are the median of the individual simulations (represented by the transparent lines). The reminding parameters in the simulations are $\{\Pen, R, \mathcal A\} = \{500, 2, 10^{-3}\}$.}
  \label{fig:random_h_and_n_vs_F}
\end{figure}

 Figure~\ref{fig:random_KA_evo} shows a contour plot of one of the simulations for each parameter set. It is only when the volumetric flux ratio $\mathcal{KA}=1$, that the primary fluid pathway occurs through the vertical fractures that connect the horizontal fractures; otherwise, the finger aligned with a fracture enters the rock matrix when the horizontal fracture ends. This qualitative behavior, observed in Figure~\ref{fig:random_KA_evo} for the evolution of the concentration, is also seen in the case of the brick shaped network.
 Figure~\ref{fig:random_h_and_n_vs_F} depicts the mixing length and the number of fingers for simulations conducted with two different values of $N = {1, 3}$. The variance of the quantitative measures between each simulation is largest for the case of the highest volumetric flux ratio ($\mathcal K\mathcal A=1$). Note that some simulations for $\mathcal K\mathcal A=1, N=3$ was ended earlier than $\hat t=100$ when the adaptivity caused a domain size domain larger than 2 500 ($h\sim 200$) due to computational demands. Based on these observations, we propose that $\mathcal{KA}=1$ serves as a threshold value, where the flow behavior is governed by the fracture geometry for values above this threshold.

A few of the generated fracture networks for the case $N=1$ did not have a horizontal fracture that crossed the initial diffusion front. In these cases, the concentration front evolved diffusively until it reached a horizontal fracture. After the first horizontal fracture was reached, the evolution continued as for the simulations with an initial fracture crossing the concentration front. In Figure~\ref{fig:random_h_and_n_vs_F} this can be best seen in the plot of the mixing length. Further, the zig-zag evolution of the mixing length is caused by the leading finger reaching the end of the connected horizontal fracture. Because the fracture networks for $N=3$ are more connected with longer chains of connected fractures and shorter distances between unconnected fractures, these jumps and plateaus in mixing length are smaller.

\subsubsection{Influence of the fracture density}
We set up a suite of simulations to investigate the impact of the fracture density for the random network. In these simulations the parameters $\{\Pen, R, \mathcal K, \mathcal A\} = \{500, 2, 10, 10^{-3}\}$ are fixed while the fracture density $N$ is varied. In this parameter regime, where the volumetric flux ratio is $\mathcal{KA} = 10^{-1}$, the variation between different random fracture networks was in Section~\ref{sec:random_F} shown to be small, and one simulation for each value of $N$ is used in this section. 

\begin{figure}[h]
  \centering%
  \def\figWidth{0.49\textwidth}%
  \def\figHeight{0.16667\textwidth}%
  \def\tikzWidth{0.9\textwidth}%
  \begin{minipage}{0.97\textwidth}
  \begin{subfigure}[b]{\figWidth}
    \centering
    \begin{tikzpicture}
      \node[inner sep=2ex, anchor=south east] (00) at (0,0) {
        \includegraphics[height=\figHeight,width=\tikzWidth]{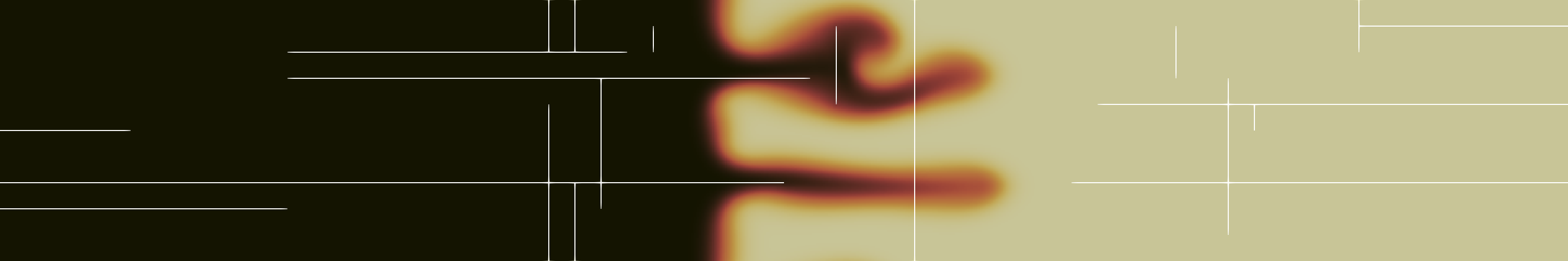}};
      \node[anchor=north, yshift=2.5ex, xshift=2.5ex] at (00.south west) {-2};
      \node[anchor=north, yshift=2.5ex] at (00.south) {1}; \node[anchor=north,
      yshift=2.5ex, xshift=-3ex] at (00.south east) {4}; \node[anchor=south,
      yshift=-1.2em] at (00.south) {$\hat x$};
      \node[anchor=east, yshift=2.5ex, xshift=2.5ex] at (00.south west) {0};
      \node[anchor=east, yshift=-2.5ex, xshift=2.5ex] at (00.north west) {1};
      \node[anchor=east, xshift=1ex] at (00.west) {$\hat y$};
    \end{tikzpicture}
    \caption{ $\hat t = 1$}
  \end{subfigure}%
  \begin{subfigure}[b]{\figWidth}
    \centering
    \begin{tikzpicture}
      \node[inner sep=2ex, anchor=south east] (00) at (0,0) {
        \includegraphics[height=\figHeight,width=\tikzWidth]{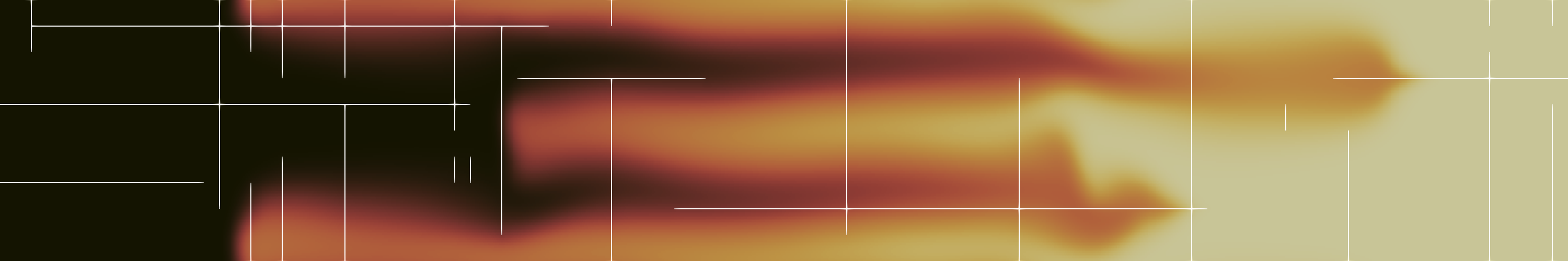}};
      \node[anchor=north, yshift=2.5ex, xshift=2.5ex] at (00.south west) {3};
      \node[anchor=north, yshift=2.5ex] at (00.south) {8}; \node[anchor=north,
      yshift=2.5ex, xshift=-3ex] at (00.south east) {13}; \node[anchor=south,
      yshift=-1.2em] at (00.south) {$\hat x$};
      \node[anchor=east, yshift=2.5ex, xshift=2.5ex] at (00.south west) {0};
      \node[anchor=east, yshift=-2.5ex, xshift=2.5ex] at (00.north west) {1};
      \node[anchor=east, xshift=1ex] at (00.west) {$\hat y$};
    \end{tikzpicture}
    \caption{ $\hat t = 8$}
  \end{subfigure}
  \begin{subfigure}[b]{\figWidth}
    \centering
    \begin{tikzpicture}
      \node[inner sep=2ex, anchor=south east] (00) at (0,0) {
        \includegraphics[height=\figHeight,width=\tikzWidth]{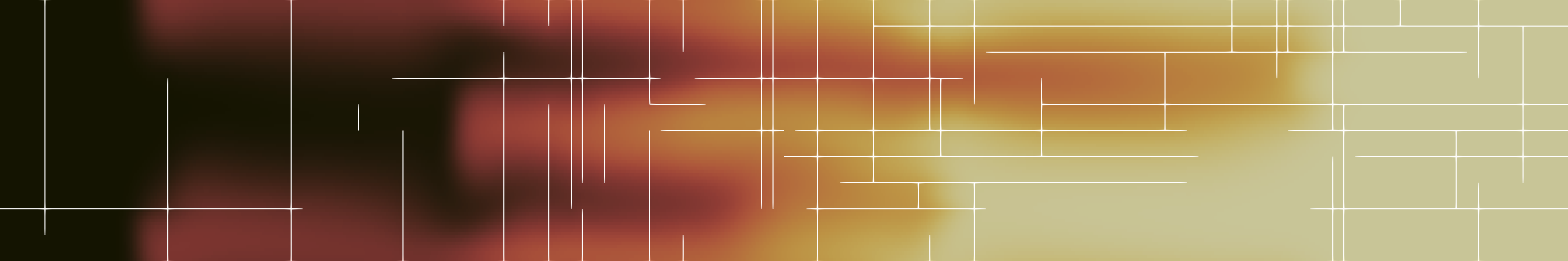}};
      \node[anchor=north, yshift=2.5ex, xshift=2.5ex] at (00.south west) {8};
      \node[anchor=north, yshift=2.5ex] at (00.south) {15}; \node[anchor=north,
      yshift=2.5ex, xshift=-3ex] at (00.south east) {22}; \node[anchor=south,
      yshift=-1.2em] at (00.south) {$\hat x$};
      \node[anchor=east, yshift=2.5ex, xshift=2.5ex] at (00.south west) {0};
      \node[anchor=east, yshift=-2.5ex, xshift=2.5ex] at (00.north west) {1};
      \node[anchor=east, xshift=1ex] at (00.west) {$\hat y$};
    \end{tikzpicture}
    \caption{$\hat t = 15$}
  \end{subfigure}%
  \begin{subfigure}[b]{\figWidth}
    \centering
    \begin{tikzpicture}
      \node[inner sep=2ex, anchor=south east] (00) at (0,0) {
        \includegraphics[height=\figHeight,width=\tikzWidth]{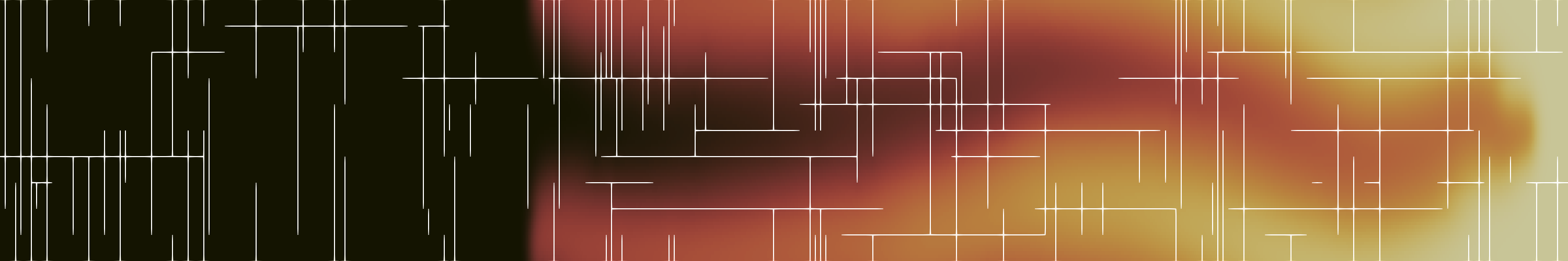}};
      \node[anchor=north, yshift=2.5ex, xshift=2.5ex] at (00.south west) {35};
      \node[anchor=north, yshift=2.5ex] at (00.south) {50}; \node[anchor=north,
      yshift=2.5ex, xshift=-3ex] at (00.south east) {65}; \node[anchor=south,
      yshift=-1.2em] at (00.south) {$\hat x$};
      \node[anchor=east, yshift=2.5ex, xshift=2.5ex] at (00.south west) {0};
      \node[anchor=east, yshift=-2.5ex, xshift=2.5ex] at (00.north west) {1};
      \node[anchor=east, xshift=1ex] at (00.west) {$\hat y$};
    \end{tikzpicture}
    \caption{$\hat t = 50$}
  \end{subfigure}%
    \end{minipage}%
  \hfill%
  \begin{minipage}[c]{.02\textwidth}
  \begin{subfigure}[b]{\textwidth}
  \includegraphics[width=\textwidth]{fig/colorbar.png}
  \end{subfigure}%
  \end{minipage}%
  \\%
  \noindent%
  \hspace{\legendWidth}
  \begin{subfigure}[T]{0.7\textwidth}
    \centering
    \includegraphics[width=1\textwidth]{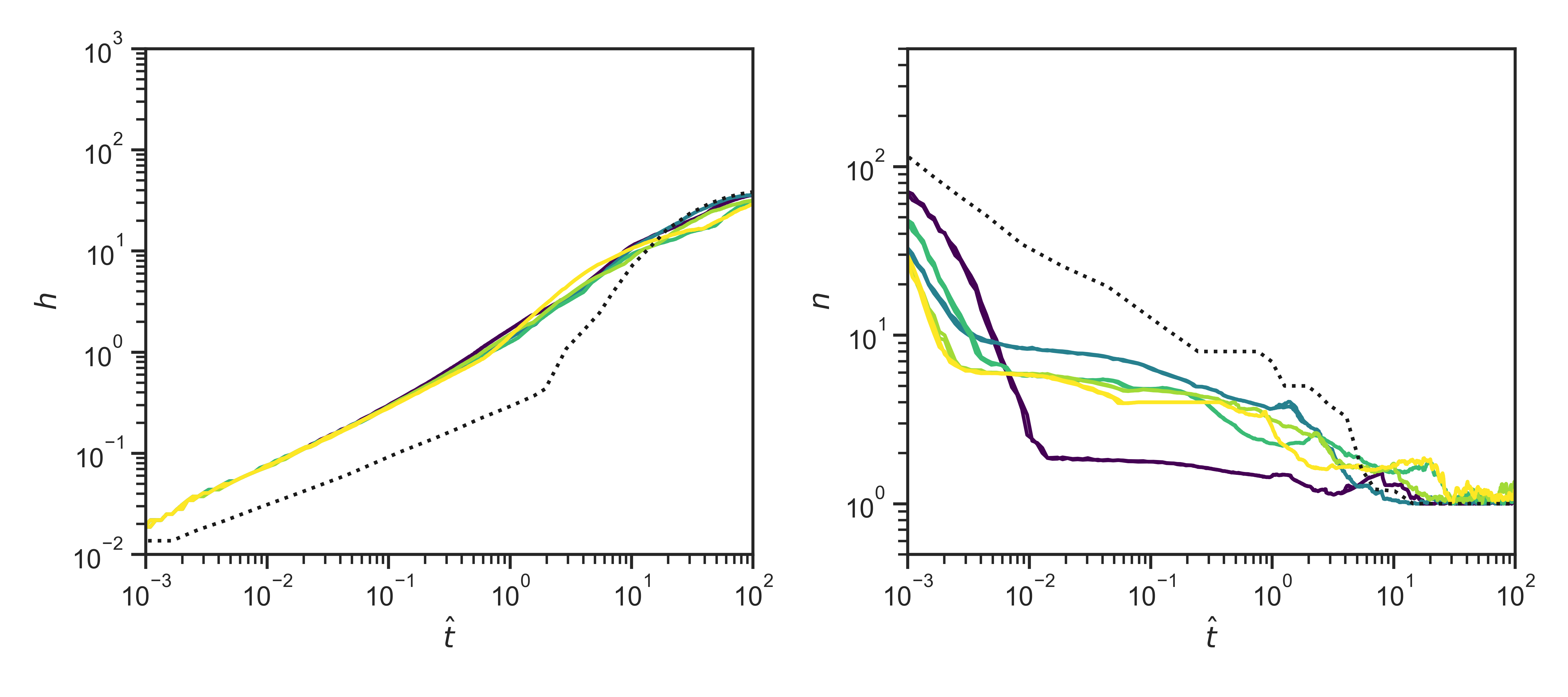}
    \caption{}
  \end{subfigure}
  \hspace{-1em}
  \begin{subfigure}[T]{\legendWidth}
    \vspace{1em}
    \includegraphics[width=1\textwidth]{fig/parallel/h_frac_vs_N_legend.png}
    \end{subfigure}
  \caption{Evolution of the concentration field for the random fracture networks. The parameters
    $\{\Pen, R, \mathcal K, \mathcal A\} = \{500, 2, 10, 10^{-3}\}$ are fixed. (a) - (d)
    Color maps of the concentration field for the case $N=3$ at different times.
    Note that the increasing aspect ratio in the figures. (e) The
    mixing length and number of fingers for different values of $N$.}
  \label{fig:random_N}
\end{figure}

Figure~\ref{fig:random_N} shows the evolution of the mixing length and the number of fingers of the simulations. In these simulations we do observe a finger that is initiated along each horizontal fracture that cross the initial concentration front at $\hat x = 0$. The initial fingers follow the horizontal fractures until they end, and we observe the viscous fingers to some extent follow new horizontal fractures they reach. However, we only observe the initiation of new fingers when the concentration front in the rock matrix reaches a new horizontal fracture for $\hat t \lesssim 1$. We believe that when the concentration front reaches a new horizontal fracture after this time, then, the viscous fingers already developed suppresses the initiation of new fingers. 

For both the brick shaped networks and the networks with parallel fractures, varying the fracture density alters the fluid flow both qualitatively and quantitatively. However, the alteration is mainly due to the repeated structure of these networks causing a stable finger to form around each horizontal fracture. Such a repeated structure is not present in the random network, and the clear transition from fingers propagating along fractures to viscous fingering across fractures that was present for the structured networks does not appear for the random networks. The randomness of the fracture network causes instabilities between the initial fingers aligned with the fractures and does not allow them to grow at equal rates as was seen in the structured networks.

\section{Discussion and concluding remarks}
This paper focuses on investigating the influence of highly permeable fractures on miscible viscous fingering in porous media using numerical simulations. We introduce three dimensionless numbers that characterize the fractures: the permeability ratio between fracture and rock matrix, the dimensionless fracture aperture, and the fracture density. Through extensive numerical simulations, we examine how these dimensionless numbers impact fluid flow for various fracture geometries. Our findings reveal that the volumetric flux ratio, which is the product of the permeability ratio and the dimensionless aperture, is the most critical parameter for describing the influence of fractures on fluid flow. This quantity represents the volume ratio of fluid flowing through the fractures compared to the fluid flowing through the rock matrix.

When fractures are present in an otherwise homogeneous domain, the early time qualitative effects observed in this paper are similar across all fracture geometries and parameter ranges. In the presence of a permeability contrast between the fractures and rock matrix, fingers aligned with the fractures always form. These fracture-induced fingers outgrow any other viscous instabilities occurring in the rock matrix, even for moderate permeability contrasts. The fracture induced fingers significantly shield the viscous fingering process within the rock matrix, thereby stabilizing the initial fingering behavior. However, the subsequent evolution of viscous fingering strongly depends on the volumetric flux ratio, leading us to classify the flow into two distinct qualitative regimes.

In the first regime, characterized by a small volumetric flux ratio ($\mathcal{KA}<10^{-1}$), the stable propagation of the fingers aligned with the fracture geometry is disrupted by viscous instabilities. These instabilities can arise from the competition between fingers aligned with different fractures or from the formation of fingers in the rock matrix between fractures. In this regime, the geometry of the fracture networks does not significantly influence the flow behavior once the viscous instabilities have emerged. Instead, the fluid behavior resembles that of a homogeneous domain, with the shutdown of viscous fingering occurring at a time scale $\hat t\sim \Pen$, where a single finger propagates independently of the fracture geometry. However, we find that the fracture geometry plays a crucial role before the onset of viscous fingering, determining when the transition to a regime dominated by viscous effects occurs.

When the volumetric flux ratio is sufficiently large, the fluid flow transitions to the second regime, where the flow is dominated by the fractures. For the fracture geometries examined in this paper, the transition to a fracture-dominated flow occurs at a volumetric flux ratio of approximately $\mathcal{KA}\sim 10^{-1}-10^0$ depending on the fracture geometry. In this regime, only channeling along the fractures is observed, while the displacement front in the rock matrix remains stable. In these cases, the flow paths are entirely determined by the fracture network. 
All investigated geometries exhibit a similar transition from unstable flow to flow dominated by the fractures as the volumetric flux ratio increases. However, the interplay between viscous instabilities and fracture geometry results in fundamentally different flow paths within the various fractured porous media. We find that the geometry of the fracture network is particularly crucial in the transition regime leading to the fracture-dominated regime. During the transition regime, the connections between fractures aligned with the flow direction by transversal fractures become important, giving rise to a intricate interactions between the fracture network geometry and fluid flow.

In sum, the examples in this paper show the complex interplay between fracture
geometry and unstable flow. The strong geometry dependency gives an important
lesson in terms of quantification and upscaling of flow and mixing: While clear
fracture flow and viscous fingering regimes can be identified, there are important
cross-over regimes where classical a priori assumptions on the fluid flow
structure fail.

\bibliography{paper}

\subsection*{Funding}
This work has been funded in part by the Norwegian Research Council grant 250223

\subsection*{Data availability}
Datasets from the simulations are available Zenodo (\href{https://doi.org/10.5281/zenodo.8020702}{https://doi.org/10.5281/zenodo.802070}). Further, the Python code that has been used in this work is archived on Zenodo (\href{https://doi.org/10.5281/zenodo.8020718}{https://doi.org/10.5281/zenodo.8020718}) and is available open source from GitHub (\href{https://github.com/rbe051/ViscFrac.git}{https://github.com/rbe051/ViscFrac.git})

\end{document}